\documentclass[10pt,floatfix,%
aps,
prl,
superscriptaddress,
amsmath,amssymb,
twocolumn,
]{revtex4-2}
\usepackage{nameref}
\usepackage[colorlinks,
 linkcolor = black,
 citecolor = red,
 urlcolor = black,
 anchorcolor = black,
 filecolor = black,
 ]{hyperref}
\usepackage{amsmath}
\usepackage{titlesec}
\usepackage{graphicx}
\usepackage{dcolumn}
\usepackage{bm}
\usepackage{rotating}
\usepackage{mathrsfs}
\usepackage[dvipsnames]{xcolor}

\titleformat{\section}[runin]
  {\normalfont\itshape}
  {}
  {0pt}
  {}
  [\textbf{---}]
\titlespacing*{\section}
  {10pt}
  {0ex}
  {0.0em}  
\begin{document}
\title{Collective Coherent Perfect Absorption in a Synthetic Photon-Phonon Lattice}

\author{Sihan Wang}
\affiliation{Laboratory of Spin Magnetic Resonance, Hefei National Research Center for Physical Sciences at the Microscale, Anhui Provincial Key Laboratory of Scientific Instrument Development and Application, University of Science and Technology of China, Hefei 230026, China}
\affiliation{Beijing Key Laboratory of Fault-Tolerant Quantum Computing, Beijing Academy of Quantum Information Sciences, Beijing 100193, China}
\affiliation{Hefei National Laboratory, University of Science and Technology of China, Hefei 230088, China}

\author{Qichun Liu}
\affiliation{Beijing Key Laboratory of Fault-Tolerant Quantum Computing, Beijing Academy of Quantum Information Sciences, Beijing 100193, China}

\author{Bo Song}
\affiliation{State key Laboratory of Artificial Microstructure and Mesoscopic Physics, School of Physics, Frontiers Science Center for Nano-optoelectronics, Peking University, Beijing 100871, China} 
\affiliation{Collaborative Innovation Center of Extreme Optics, Shanxi University, Taiyuan, Shanxi 030006, China}

\author{Qiongyi He}
\affiliation{State key Laboratory of Artificial Microstructure and Mesoscopic Physics, School of Physics, Frontiers Science Center for Nano-optoelectronics, Peking University, Beijing 100871, China} 
\affiliation{Collaborative Innovation Center of Extreme Optics, Shanxi University, Taiyuan, Shanxi 030006, China}

\author{Jingwei Zhou}
\email{zhoujw@ustc.edu.cn}
\affiliation{Laboratory of Spin Magnetic Resonance, Hefei National Research Center for Physical Sciences at the Microscale, Anhui Provincial Key Laboratory of Scientific Instrument Development and Application, University of Science and Technology of China, Hefei 230026, China}
\affiliation{Hefei National Laboratory, University of Science and Technology of China, Hefei 230088, China}

\author{Yulong Liu}
\email{liuyl@baqis.ac.cn}
\affiliation{Beijing Key Laboratory of Fault-Tolerant Quantum Computing, Beijing Academy of Quantum Information Sciences, Beijing 100193, China}

\date{\today}
\begin{abstract}

Coherent perfect absorption (CPA) has emerged as a powerful paradigm for controlling classical and quantum light, and has been demonstrated across a broad range of physical platforms. CPA realized in optomechanics relies on interference between the input field and the mechanically scattered field, but is intrinsically confined to the weak-cooperativity regime, resulting in a narrow absorption bandwidth and the mechanical mode remains thermally occupied. Here, we experimentally demonstrate collective interference-induced CPA in a synthetic photon--phonon lattice. By harnessing cavity-reservoir-mediated interactions among Floquet lattice sites, collective interference shifts the CPA condition deep into the high-cooperativity regime. This enables CPA to coexist with ground-state cooling of the mechanical oscillator, together with a broadened non-Lorentzian absorption lineshape and a singular group-delay response. Our results identify collective interference as a route to quantum-compatible perfect absorption and long-lived quantum storage.
\end{abstract}

\maketitle
\section*{Introduction}
Coherent perfect absorption (CPA) is the complete absorption of incoming coherent waves, enabled by the interplay between coherent interference and dissipation~\cite{baranov2017Coherent}. Since its introduction for light interacting with absorbing scatterers~\cite{chong2010Coherent,wan2011TimeReversed}, CPA has been demonstrated in a wide variety of physical platforms, including optical resonators~\cite{wang2021Coherent,horner2024Coherent,xue2025DualColor,galiffi2026Optical}, disorder medium~\cite{horodynski2022Antireflection,noh2012Perfecta}, metamaterials~\cite{roger2015Coherent}, acoustic systems~\cite{ma2014Acoustic,xia2025Observation}, parity-time symmetric structures~\cite{sun2014Experimentala,feng2017NonHermitian}, exciton--polariton devices~\cite{noh2012Perfect,zanotto2014Perfect,zhang2017Observationa} and optomechanical (OM) systems~\cite{weis2010Optomechanically,massel2011Microwave, safavi-naeini2011Electromagnetically,hocke2012Electromechanically,zhou2013Slowing}. 
Moreover, CPA promise practical applications such as all-optical control and sensing~\cite{pendry2006Controlling,rechtsman2017Optical,liu2010Infrared}, while also allows for generalized quantum state engineering~\cite{vetlugin2021Coherent}, including photon anticoalescence~\cite{vest2017Anticoalescence}, single-photon absorption~\cite{roger2015Coherent} and the manipulation of non-classical states~\cite{roger2016Coherent,altuzarra2017Coherent,hernandez2022Quantum}. Extending CPA to quantum systems is therefore of fundamental importance for coherent quantum control and information processing~\cite{baranov2017Coherent,mullers2018Coherent,lai2024Roomtemperature}.

Among various quantum platforms, OM systems provide a versatile platform for coherently manipulating light and macroscopic quantum states~\cite{aspelmeyer2014Cavityb,pirkkalainen2013Hybrid,oconnell2010Quantum}. Optomechanical CPA (OM-CPA) has been realized through interference between directly reflected and mechanically scattered fields~\cite{safavi-naeini2011Electromagnetically,hocke2012Electromechanically,zhou2013Slowing}, giving rise to a characteristic phase singularity and divergent group delays~\cite{liu2021Optomechanical,liu2025Degeneracybreaking}. However, conventional OM-CPA is fundamentally restricted to the weak-cooperativity regime, where the mechanical oscillator remains thermally occupied and the absorption bandwidth is intrinsically limited by the mechanical linewidth~\cite{teufel2011Sidebanda, aspelmeyer2014Cavityb}. These intrinsic constraints prevent operation in the quantum regime of mechanical motion and substantially limit the potential of OM-CPA as a coherent quantum memory~\cite{palomaki2013Coherent,liu2023Coherenta,bozkurt2025Mechanical}.
\par

\begin{figure*}[t]
\includegraphics[scale=1.0]{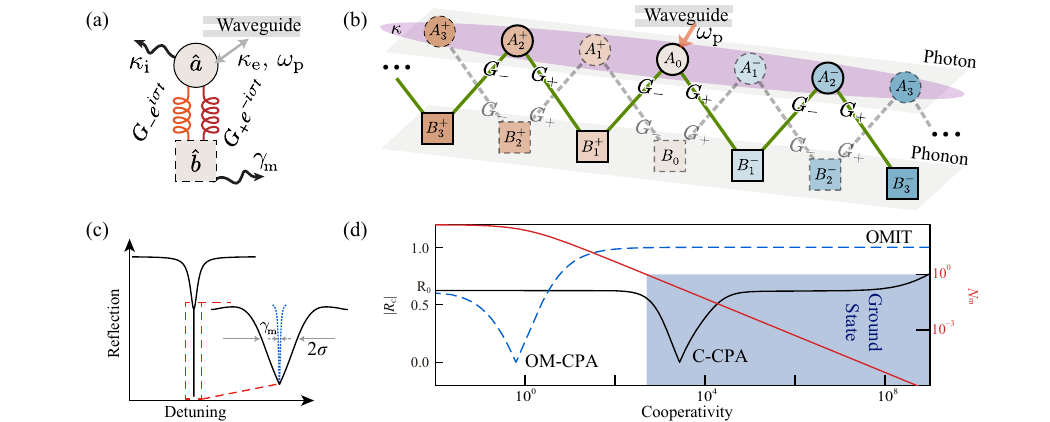}
\caption{\label{fig1}
(a) Schematic of the Floquet-engineered optomechanical coupling. 
(b) Equivalent multimode coupling chain, where $A_n^{\pm}$ and $B_n^{\pm}$ denote the photonic and phononic Floquet sites, respectively. The purple shaded region denotes the common cavity thermal reservoir with decay rate $\kappa$.
(c) Typical C--CPA curve in the reflection spectrum, with the inset highlighting the detailed lineshape determined by the offset $\sigma$. The blue dashed curve denotes the corresponding OM-CPA response for comparison.
(d) Dependence of $|R_\mathrm{c}|$ and the effective phonon number $N_\text{m}$ on the cooperativity.
The black curve denotes collectively photon-phonon interference induced absorption, whereas the blue dashed line corresponds to traditional two mode interference induced absorption and transparency.
The red curve illustrates the evolution of the effective phonon number, taking an initial occupancy of 500 phonons as an example, and the shaded region represents the ground-state regime ($N_\text{m}<1$).
}
\end{figure*}

In this Letter, we construct a synthetic photon–phonon lattice via designing time‑dependent optomechanical couplings and experimentally demonstrate collective interference‑induced coherent perfect absorption (C‑CPA) in the high‑cooperativity regime. This effect arises from long‑range interactions and cooperative interference among multiple synthetic lattice sites, mediated by the shared reservoir.
As a consequence, we uncover a group-delay divergence singularity accompanying C-CPA and achieve a group delay of up to $\mathrm{10~s}$, while the mechanical oscillator enters its ground state. It further enables a non-Lorentzian absorption profile with a bandwidth that exceeds the intrinsic mechanical linewidth, a broadening of three orders of magnitude.
Our results extend the operating regime of traditional OM‑CPA into the quantum domain for the first time, establishing a pathway toward thermal‑noise‑free and long‑lived on‑chip OM quantum storage.

\section*{Photon-phonon chains and C-CPA}
In a standard optomechanical system, a single coherent pump at frequency $\omega_{\mathrm{L}}$ induces a linearized optomechanical coupling $G = g_{0}\sqrt{n_{\mathrm{c}}}$, where $g_{0}$ is the single-photon coupling rate and $n_{\mathrm{c}}$ denotes the intracavity photon number~\cite{aspelmeyer2014Cavityb}. 
As illustrated in Fig.~\ref{fig1}(a), we establish a time-dependent interaction by applying two pump tones separated by a small frequency offset $\sigma$. 
Their detunings are defined as $\Delta_{\pm} = \Delta \pm \sigma$, where the central detuning $\Delta = \omega_{\mathrm{L}} - \omega_{\mathrm{c}}$ is set to the red sideband of the cavity resonance $\omega_{\text{c}}$ at the mechanical frequency $\omega_{\text{m}}$ (i.e., $\Delta = -\omega_{\text{m}}$). The cavity and mechanical modes are described by the bosonic operators $\hat{a}$ and $\hat{b}$, respectively.

\begin{figure}
\includegraphics[scale=1.0]{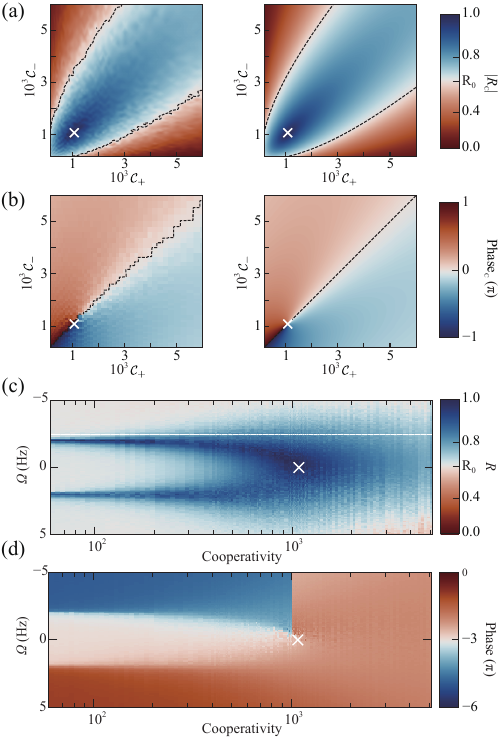}
\caption{\label{fig2}
(a) Measured (left) and simulated (right) resonance reflection amplitude $|R_{\mathrm{c}}|$ at resonance as a function of cooperativities $\mathcal{C}_+$ and $\mathcal{C}_-$. Black dashed line denotes the baseline reflection $|R_{\mathrm{c}}|=R_0$. 
(b) Measured (left) and simulated (right) wrapped reflection phase $\varphi_{\mathrm{c}}$ as a function of $\mathcal{C}_+$ and $\mathcal{C}_-$. Black dashed lines indicate $\varphi_{\mathrm{c}}=0$ and $\varphi_{\mathrm{c}}=\pm\pi$.
(c) The reflection amplitude as a function of cooperativity, showing the continuous variation and spectral dips merging. The white dotted line marks the location of dips.
(d) The reflection phase as a function of cooperativity. The pronounced phase step observed prior to C-CPA collapses into a degeneracy at the C-CPA point. White crosses mark the C-CPA locations.
}
\end{figure}

The resulting time-dependent interaction Hamiltonian takes the form
$
\mathcal{H}_\text{int}(t)
=
\hbar G_+ e^{-i\sigma t}\hat a \hat b^\dagger
+
\hbar G_- e^{i\sigma t}\hat a \hat b^\dagger
+ \text{H.c.},
$
where $G_{\pm}$ denotes the parametric coupling rates associated with the two pump tones at detunings $\Delta_{\pm}$. The frequency offset $\sigma$ acts as an effective coupling modulation frequency.
Owing to the temporal periodicity, the system admits a Floquet description by expanding the fields in the basis $\{e^{i n\sigma t}\}$ as $\langle \hat{a} \rangle = \sum_n A_n e^{-i n \sigma t}$ and $\langle \hat{b} \rangle = \sum_n B_n e^{-i n \sigma t}$~\cite{bukov2015Universal,wang2026SI} .

Including the cavity decay rate $\kappa$ and the mechanical damping rate $\gamma_\mathrm{m}$, the Floquet system is mapped onto a synthetic frequency lattice described by the effective non-Hermitian Hamiltonian
\begin{equation}
\begin{aligned}
H_{\mathrm{lat}}
&=
\sum_n
\Big[
\epsilon_A[n] A_n^\dagger A_n
+
\epsilon_B[n] B_n^\dagger B_n
\Big]\\
&+
\sum_n
\Big[
t_1 A_n^\dagger B_{n-1}
+
t_2 A_n^\dagger B_{n+1}
+\mathrm{H.c.}
\Big],
\end{aligned}
\end{equation}
with the on-site energies $\epsilon_A[n]=n\sigma-i\frac{\kappa}{2},\,\epsilon_B[n]=n\sigma-i\frac{\gamma_{\mathrm m}}{2}$, and the hopping amplitudes $t_1=iG_+,\,t_2=iG_-$.

Here, the Floquet index $n$ introduces a linear on-site potential, while optical and mechanical dissipation render the lattice intrinsically non-Hermitian. Consequently, the driven optomechanical system is equivalent to a bipartite Wannier--Stark lattice in synthetic frequency space, where photonic and phononic Floquet sidebands form two interleaved sublattices connected through photon--phonon hopping, as illustrated in Fig.~\ref{fig1}(b).

The response of the synthetic lattice to a weak probe field with amplitude $\alpha_{\mathrm p}$ and frequency $\omega_{\mathrm p}$ is determined by the resolvent $(\Omega I-H_{\mathrm{lat}})\Psi=F$, where $\Omega=\omega_{\mathrm p}-\omega_{\mathrm c}$ denotes the probe detuning from the cavity resonance, $\Psi=(\cdots,A_n,B_n,\cdots)^{\mathrm T}$ is the Floquet-state vector, and the driving vector $F=\sqrt{\kappa_{\mathrm e}}\alpha_{\mathrm p}(\cdots,0,1,0,\cdots)^{\mathrm T}$ indicates that only the central photonic site $A_0$ is directly excited by the probe.

Under the hierarchy of dissipative rates $\gamma_\mathrm{m} \ll \sigma \ll \kappa$, cavity dissipation effectively defines a common thermal reservoir shared by all Floquet lattice sites. As a result, energy redistributed across the entire lattice undergoes collective interference, such that the reflection coefficient $R$ is governed by the coherent superposition of contributions from all lattice sites within the input--output formalism:
\begin{equation}
R = 1 - 
\frac{\eta \kappa}{
\mathcal{I}[0]
- \frac{\mathcal{J}[1]}{\mathcal{I}[2] - \frac{\mathcal{J}[3]}{\mathcal{I}[4] - \cdots}}
- \frac{\mathcal{J}[-1]}{\mathcal{I}[-2] - \frac{\mathcal{J}[-3]}{\mathcal{I}[-4] - \cdots}}
}.
\end{equation}
Here, we define the optical and mechanical susceptibilities of the $n$-th Floquet sideband as $\chi_\mathrm{c}[n]=[-i(\Omega + n\sigma)+\kappa/2]^{-1}$ and $\chi_\mathrm{m}[n] = [-i(\Omega + n\sigma) + \gamma_\mathrm{m}/2]^{-1}$, respectively. And the $n$-th isolated contribution is $\mathcal{I}[n] = 1/\chi_\text{c}[n] + \chi_\text{m}[n-1]|G_-|^2 + \chi_\text{m}[n+1]|G_+|^2$ and the overlapping interference term is $\mathcal{J}[n] = |G_+|^2|G_-|^2 |\chi_\text{m}[n]|^2$. The parameter $\eta \equiv \kappa_e/\kappa$ denotes the external coupling ratio. And we define $R_0 = |2\eta - 1|$ as the bare cavity reflection baseline in the absence of collective interference.
In particular, it admits a typical C-CPA condition at resonance ($\Omega = 0$), where the reflection coefficient $R_\mathrm{c}$ vanishes due to the interference from all Floquet lattice sites.

This mechanism yields two transformative improvements over conventional OM-CPA. First, C-CPA exhibits a non-Lorentzian absorption profile with an effective linewidth $\sim 2\sigma$ [Fig.~\ref{fig1}(c)], thereby decoupling the bandwidth from the narrow mechanical dissipation and enhancing the absorption window beyond the $\eta\gamma_{\rm m}$ limit of OM-CPA by a factor of $\sigma/\gamma_{\rm m}$.
Second, the C-CPA condition can be remarkably shifted into the high-cooperativity regime ($\mathcal{C} \gg 1$), where red-sideband driving simultaneously cools the mechanical mode and suppresses its thermal occupation [Fig.~\ref{fig1}(d)]. This contrasts with conventional OM-CPA, which is restricted to $\mathcal{C}<1$ and remains thermally populated~\cite{weis2010Optomechanically,liu2021Optomechanical}, whereas at large cooperativity ($\mathcal{C}\gg1$) the system instead enters the regime of optomechanically induced transparency (OMIT).
Here, the $\mathcal{C}_\pm = 4|G_\pm|^2/(\kappa\gamma_\mathrm{m})$ represent the optomechanical cooperativities.

\section*{Critical condition and mode hybridization}
Owing to the Jacobi-type recursive continued-fraction structure of the reflection coefficient $R$, closed-form expressions for its roots are generally intractable. We therefore focus on the resonant condition ($\Omega=0$), where C-CPA corresponds to complete cancellation of the output field. This requires operation in the overcoupled regime ($\eta>1/2$), which ensures efficient excitation of the synthetic lattice. For balanced sideband couplings, $\mathcal{C}_{+}=\mathcal{C}_{-}=\mathcal{C}$, the C-CPA condition is approximately given by
$
\mathcal{C}\gamma_{\mathrm m}\approx(2\eta-1)\,2\sigma,
$
which captures the balance between cavity-mediated collective dissipation and energy redistribution across the synthetic photon--phonon lattice. Consequently, the critical cooperativity $\mathcal{C}_{\mathrm{CPA}}$ can be continuously tuned through the modulation frequency $\sigma$.

For the experiment, we realize the OM system using a superconducting microwave resonator coupled to a suspended silicon carbide (SiC) membrane. The device is mounted on the mixing chamber stage of a dilution refrigerator at $10~\mathrm{mK}$. Device fabrication follows the procedure reported in Ref.~\cite{liu2025Degeneracybreaking}. Detailed measurement setup and device parameters can be found in Section~III of Ref.~\cite{wang2026SI}.
The device operates in the overcoupled regime with an external coupling ratio $\eta = 0.8125$, corresponding to a bare cavity reflection baseline $R_0 = |2\eta - 1| = 0.625$. The cavity and mechanical decay rates are independently calibrated as $\kappa / 2\pi = 158.8~\mathrm{kHz}$ and $\gamma_\mathrm{m} / 2\pi = 3.4~\mathrm{mHz}$, respectively.

We fix the modulation frequency at $\sigma/2\pi=2~\mathrm{Hz}$ and map the resonant reflection amplitude $|R_{\mathrm c}|$ and phase $\varphi_{\mathrm c}$ as functions of $\mathcal{C}_{+}$ and $\mathcal{C}_{-}$. C-CPA emerges at a distinct point along the diagonal $\mathcal{C}_{+}=\mathcal{C}_{-}$ in the parameter space [Figs.~\ref{fig2}(a)].
Along this diagonal, a sharp phase boundary is observed, accompanied by a $\pi$-phase transition in $\varphi_\mathrm{c}$ across the C-CPA point [Figs.~\ref{fig2}(b)]. The measured evolution agrees well with theoretical predictions, with minor deviations attributed to residual parameter fluctuations. For $\sigma / 2\pi = 2~\mathrm{Hz}$, we obtain $\mathcal{C}_\mathrm{CPA} = 1083$.

The corresponding evolution of the full reflection spectrum under symmetric couplings $\mathcal{C}_{+} = \mathcal{C}_{-}$ is shown in Figs.~\ref{fig2}(c,d). In the weak-cooperativity regime, two well-separated absorption resonances are associated with the neighboring synthetic sites $B_{-1}$ and $B_{+1}$. Increasing the cooperativity enhances hybridization among the Floquet sidebands, causing these resonances to overlap and merge into a single collective absorption centered at $\Omega=0$, where C-CPA is established. Concurrently, the reflection phase evolves from a three-step profile into a sharply varying singular response, revealing the crossover from conventional few-mode interference to collective interference across the synthetic lattice.

\section*{Spectrum evolution and phase singularity}
\begin{figure}
\includegraphics{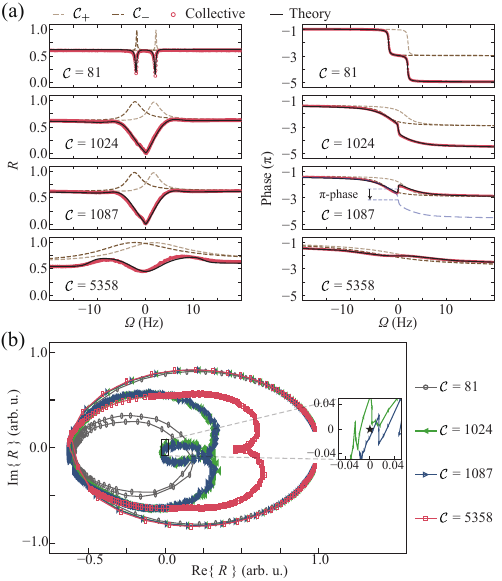}
\caption{
\label{fig3}
(a) Four characteristic regions of the reflection spectrum and phase. Red circles represent the experimental results of collective interference, the black solid line represents the theoretical results, and the dashed line indicates the experimental results for the effect of $\mathcal{C}_\pm$ acting individually. 
This corresponds to $\sigma / 2\pi = 2 \, \text{Hz}$. In the subpanel corresponding to $\mathcal{C} = 1087$, the blue dashed line highlights the $\pi$-phase transition at $\Omega=0$.
(b) Complex-plane plots of the real and imaginary parts of $R$. The winding number abruptly changes from $2$ to $1$ after the system passes through the CPA singularity, where non-closed cavity trajectories are neglected. An inset highlights the change of the winding number as $\mathcal{C}$ crosses the critical value $\mathcal{C}_{\text{CPA}}$. The origin $(0,0)$ is marked by a black pentagram.
}
\end{figure}
Four representative operating regimes at $\sigma / 2\pi = 2~\mathrm{Hz}$ are shown in Fig.~\ref{fig3}(a). 
In the regime $1 \ll \mathcal{C} \ll \mathcal{C}_\mathrm{CPA}$, collective interference gives rise to two symmetric absorption dips in the reflection spectrum, accompanied by discrete phase steps, as illustrated at $\mathcal{C} \approx 81$. This behavior cannot be described as a simple superposition of two independent optomechanically induced absorption processes. Indeed, for $\mathcal{C} > 1$, each individual pathway (i.e., with only $\mathcal{C}_{+}$ or $\mathcal{C}_{-}$ present) already operates in the OMIT regime, highlighting the intrinsically collective origin of the observed features.

As the cooperativity $\mathcal{C}$ approaches and exceeds the critical value $\mathcal{C}_\mathrm{CPA}$, a sharp absorption dip with vanishing reflectivity emerges at $\Omega = 0$, constituting a hallmark of CPA. At the C-CPA condition, the reflected probe is suppressed by $36.81\pm4.15~\mathrm{dB}$, corresponding to a residual power reflection coefficient $|R|^2\simeq2\times10^{-4}$.
The quoted uncertainty is obtained from the mean and standard deviation of long-time fluctuations. Concurrently, the merging of Floquet sidebands leads to a non-Lorentzian lineshape, with an effective linewidth of $2\sigma$, viz, $4~\mathrm{Hz}$. This represents an enhancement of the absorption bandwidth over conventional OM-CPA by a factor of $2\sigma / \gamma_\mathrm{m} \sim 10^3$, which can be further increased by tuning the modulation frequency $\sigma$.

Across the critical point, the phase evolution undergoes a qualitative transition: below $\mathcal{C}_\mathrm{CPA}$, it accumulates a $4\pi$ variation over the frequency interval $\Omega \in [-10, 10]~\mathrm{Hz}$ with a negative slope at resonance, whereas above it the phase becomes folded, displaying a reduced $2\pi$ variation and a positive slope. This abrupt transition is captured by a change in the winding number of $R$ in the complex plane, which decreases from $2$ to $1$ as the system crosses the C-CPA condition, as shown in Fig.~\ref{fig3}(b). 
For larger cooperativity, higher-order Floquet sites become increasingly significant, leading to a further broadening of the absorption dip and a smoother phase response.
\section*{Infinite group delay in quantum regime}

\begin{figure}
\includegraphics{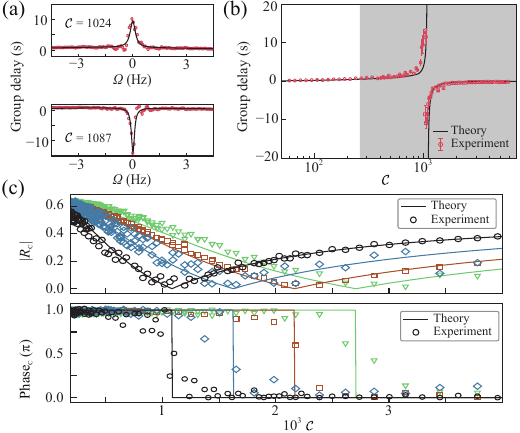}
\caption{\label{fig4}
(a) Measured group delay $\tau(\Omega)$ as a function of frequency detuning $\Omega$ for two representative cooperativities near the C-CPA. 
(b) Group delay at the cavity resonance as a function of cooperativity $\mathcal{C}$. The gray-shaded region indicates the ground state regime with $N_\text{m} < 1$. Error bars represent the standard error of the mean from repeated measurements.
(c) Dependence of amplitude $|R_\text{c}|$ and phase $\varphi_\text{c}$ on the cooperativity $\mathcal{C}$ for different modulation frequencies $\sigma$. Results are shown for $\sigma/2\pi = 2$ (black circles), 3 (blue diamonds), 4 (orange squares), and 5 (green triangles) Hz.
}
\end{figure}

The reflection phase $\varphi_\mathrm{c}$ exhibits a $\pi$-phase transition, giving rise to a steep phase gradient near resonance. The corresponding group delay is defined as $\tau = -\partial \varphi / \partial \Omega$. Consequently, the singular phase response at $\mathcal{C}_\mathrm{CPA}$ leads to a non-analytic behavior in the phase derivative, which in principle results in a divergent group delay.

Fig.~\ref{fig4}(a) presents the measured group delay for two cooperativities straddling $\mathcal{C}_\mathrm{CPA}$, showing a pronounced peak near $\Omega = 0$. For $\mathcal{C} < \mathcal{C}_\mathrm{CPA}$, a large positive group delay indicates slow-light behavior, whereas for $\mathcal{C} > \mathcal{C}_\mathrm{CPA}$, the delay becomes strongly negative, corresponding to fast-light propagation. While the group delay is expected to diverge at the singular point, device fluctuations and measurement instabilities limit the observed values to the order of $10~\mathrm{s}$. The group delay at resonance, $\tau_\mathrm{c}$, as a function of cooperativity is shown in Fig.~\ref{fig4}(b), where a clear sign reversal is observed across $\mathcal{C}_\mathrm{CPA}$. As $\mathcal{C}$ approaches $\mathcal{C}_\mathrm{CPA}$ from either side, the delay (or advance) exhibits a pronounced divergence.

Furthermore, the realization of C-CPA in our system occurs in the high-cooperativity regime, while the red-sideband pump tones simultaneously enable cooling of the mechanical oscillator.
Neglecting photon number fluctuations in the input fields, the effective phonon occupation is given by
$N_\mathrm{m} = N_\mathrm{T}/(\mathcal{C}_{+} + \mathcal{C}_{-} + 1)$,
where $N_\mathrm{T}$ denotes the thermal phonon occupation set by the bath temperature.
For the mechanical oscillator in our setup, with $\omega_\mathrm{m}/2\pi = 1.18~\mathrm{MHz}$ and $T = 30~\mathrm{mK}$, achieving $N_\mathrm{m} < 1$ requires a total cooperativity $\mathcal{C}_{+} + \mathcal{C}_{-} \gtrsim 530$.
The critical cooperativity for realizing C-CPA in our system, $\mathcal{C}_{+} + \mathcal{C}_{-} = 2\mathcal{C}_\mathrm{CPA} \approx 2166$, lies well above this threshold, indicating that C-CPA is naturally compatible with ground-state cooling (see Section~IV of Ref.~\cite{wang2026SI}).

In contrast, the critical cooperativity for conventional OM-CPA is fixed and typically constrained to $\mathcal{C} < 1$. In the large-cooperativity regime ($\mathcal{C} \gg 1$), the system instead enters the OMIT regime, where the phase response becomes trivial. As a result, it is incompatible with the simultaneous realization of a divergent group delay, characteristic of CPA, and ground-state operation of the mechanical oscillator.

The critical cooperativity $\mathcal{C}_{\mathrm{CPA}}$ depends sensitively on the modulation frequency $\sigma$. As shown in Fig.~\ref{fig4}(c), for $\sigma/2\pi = 2, 3, 4,$ and $5~\mathrm{Hz}$, the value of $\mathcal{C}_{\mathrm{CPA}}$ corresponding to vanishing resonant reflection and the $\pi$-phase transition increases monotonically with $\sigma$. This behavior highlights the high degree of tunability inherent to C-CPA, enabling its realization at a controllable cooperativity by adjusting $\sigma$ within experimental constraints. In experiment, the phase evolution exhibits a finite slope rather than an ideal step, which can be attributed to a residual imbalance between $\mathcal{C}_+$ and $\mathcal{C}_-$.

\section*{Conclusions}

We have demonstrated collective interference-induced coherent perfect absorption in a Floquet-engineered synthetic photon--phonon lattice. Collective interference shifts the CPA condition from $\mathcal{C}<1$ to the order of $10^3$ and thereby overcoming the incompatibility between OM-CPA and quantum ground-state cooling. The resulting C-CPA exhibits a non-Lorentzian absorption spectrum with a Hz-scale linewidth and a phase singularity that produces group delays on the order of $10~\mathrm{s}$. These features open new avenues for OM systems in implementing ultra-long optical storage and dynamically reconfigurable slow and fast light in hybrid quantum architectures~\cite{mirhosseini2020Superconducting,pirkkalainen2013Hybrid,rodrigues2019Coupling,zoepfl2020SinglePhoton}. More broadly, the synthetic photon--phonon lattice provides a versatile platform for exploring many-body dynamics~\cite{fazio2025ManyBody,diehl2010Dynamical}, coherent frequency comb~\cite{wang2026Pumpthresholdfree}, topological photonics\cite{youssefi2022Topological,kitaev2001unpaired}, and quantum simulation in synthetic dimensions~\cite{lindner2011Floquet}, with potential extensions to paradigmatic lattice models and diverse cavity-based quantum platforms~\cite{walther2006Cavity,zhao2021PhaseControlled,blais2021Circuit,ozdemir2019Parity}.

\section*{Acknowledgments}
Y.~L.~acknowledges the support of National Natural Science Foundation of China (Nos.~12374325 and ~92365210), Beijing Natural Science Foundation (Z240007), CAST Young Elite Scientists Sponsorship Program (Grant No.~2023QNRC001). This work was also supported by the Quantum Science and Technology-National Science and Technology Major Project (Grant No. 2021ZD0302200).

\section*{Data availability}
There are no publicly available research data or software supporting this manuscript.
Requests for further information or data should be sent
to the authors.

\bibliography{Main}

\begin{thebibliography}{56}%
\makeatletter
\providecommand \@ifxundefined [1]{%
 \@ifx{#1\undefined}
}%
\providecommand \@ifnum [1]{%
 \ifnum #1\expandafter \@firstoftwo
 \else \expandafter \@secondoftwo
 \fi
}%
\providecommand \@ifx [1]{%
 \ifx #1\expandafter \@firstoftwo
 \else \expandafter \@secondoftwo
 \fi
}%
\providecommand \natexlab [1]{#1}%
\providecommand \enquote  [1]{``#1''}%
\providecommand \bibnamefont  [1]{#1}%
\providecommand \bibfnamefont [1]{#1}%
\providecommand \citenamefont [1]{#1}%
\providecommand \href@noop [0]{\@secondoftwo}%
\providecommand \href [0]{\begingroup \@sanitize@url \@href}%
\providecommand \@href[1]{\@@startlink{#1}\@@href}%
\providecommand \@@href[1]{\endgroup#1\@@endlink}%
\providecommand \@sanitize@url [0]{\catcode `\\12\catcode `\$12\catcode `\&12\catcode `\#12\catcode `\^12\catcode `\_12\catcode `\%12\relax}%
\providecommand \@@startlink[1]{}%
\providecommand \@@endlink[0]{}%
\providecommand \url  [0]{\begingroup\@sanitize@url \@url }%
\providecommand \@url [1]{\endgroup\@href {#1}{\urlprefix }}%
\providecommand \urlprefix  [0]{URL }%
\providecommand \Eprint [0]{\href }%
\providecommand \doibase [0]{https://doi.org/}%
\providecommand \selectlanguage [0]{\@gobble}%
\providecommand \bibinfo  [0]{\@secondoftwo}%
\providecommand \bibfield  [0]{\@secondoftwo}%
\providecommand \translation [1]{[#1]}%
\providecommand \BibitemOpen [0]{}%
\providecommand \bibitemStop [0]{}%
\providecommand \bibitemNoStop [0]{.\EOS\space}%
\providecommand \EOS [0]{\spacefactor3000\relax}%
\providecommand \BibitemShut  [1]{\csname bibitem#1\endcsname}%
\let\auto@bib@innerbib\@empty
\bibitem [{\citenamefont {Baranov}\ \emph {et~al.}(2017)\citenamefont {Baranov}, \citenamefont {Krasnok}, \citenamefont {Shegai}, \citenamefont {Al{\`u}},\ and\ \citenamefont {Chong}}]{baranov2017Coherent}%
  \BibitemOpen
  \bibfield  {author} {\bibinfo {author} {\bibfnamefont {D.~G.}\ \bibnamefont {Baranov}}, \bibinfo {author} {\bibfnamefont {A.}~\bibnamefont {Krasnok}}, \bibinfo {author} {\bibfnamefont {T.}~\bibnamefont {Shegai}}, \bibinfo {author} {\bibfnamefont {A.}~\bibnamefont {Al{\`u}}},\ and\ \bibinfo {author} {\bibfnamefont {Y.}~\bibnamefont {Chong}},\ }\bibfield  {title} {\bibinfo {title} {Coherent perfect absorbers: Linear control of light with light},\ }\href {https://doi.org/10.1038/natrevmats.2017.64} {\bibfield  {journal} {\bibinfo  {journal} {Nat. Rev. Mater}\ }\textbf {\bibinfo {volume} {2}},\ \bibinfo {pages} {17064} (\bibinfo {year} {2017})}\BibitemShut {NoStop}%
\bibitem [{\citenamefont {Chong}\ \emph {et~al.}(2010)\citenamefont {Chong}, \citenamefont {Ge}, \citenamefont {Cao},\ and\ \citenamefont {Stone}}]{chong2010Coherent}%
  \BibitemOpen
  \bibfield  {author} {\bibinfo {author} {\bibfnamefont {Y.~D.}\ \bibnamefont {Chong}}, \bibinfo {author} {\bibfnamefont {L.}~\bibnamefont {Ge}}, \bibinfo {author} {\bibfnamefont {H.}~\bibnamefont {Cao}},\ and\ \bibinfo {author} {\bibfnamefont {A.~D.}\ \bibnamefont {Stone}},\ }\bibfield  {title} {\bibinfo {title} {Coherent {{Perfect Absorbers}}: {{Time-Reversed Lasers}}},\ }\href {https://doi.org/10.1103/PhysRevLett.105.053901} {\bibfield  {journal} {\bibinfo  {journal} {Phys. Rev. Lett.}\ }\textbf {\bibinfo {volume} {105}},\ \bibinfo {pages} {053901} (\bibinfo {year} {2010})}\BibitemShut {NoStop}%
\bibitem [{\citenamefont {Wan}\ \emph {et~al.}(2011)\citenamefont {Wan}, \citenamefont {Chong}, \citenamefont {Ge}, \citenamefont {Noh}, \citenamefont {Stone},\ and\ \citenamefont {Cao}}]{wan2011TimeReversed}%
  \BibitemOpen
  \bibfield  {author} {\bibinfo {author} {\bibfnamefont {W.}~\bibnamefont {Wan}}, \bibinfo {author} {\bibfnamefont {Y.}~\bibnamefont {Chong}}, \bibinfo {author} {\bibfnamefont {L.}~\bibnamefont {Ge}}, \bibinfo {author} {\bibfnamefont {H.}~\bibnamefont {Noh}}, \bibinfo {author} {\bibfnamefont {A.~D.}\ \bibnamefont {Stone}},\ and\ \bibinfo {author} {\bibfnamefont {H.}~\bibnamefont {Cao}},\ }\bibfield  {title} {\bibinfo {title} {Time-{{Reversed Lasing}} and {{Interferometric Control}} of {{Absorption}}},\ }\href {https://doi.org/10.1126/science.1200735} {\bibfield  {journal} {\bibinfo  {journal} {Science}\ }\textbf {\bibinfo {volume} {331}},\ \bibinfo {pages} {889} (\bibinfo {year} {2011})}\BibitemShut {NoStop}%
\bibitem [{\citenamefont {Wang}\ \emph {et~al.}(2021)\citenamefont {Wang}, \citenamefont {Sweeney}, \citenamefont {Stone},\ and\ \citenamefont {Yang}}]{wang2021Coherent}%
  \BibitemOpen
  \bibfield  {author} {\bibinfo {author} {\bibfnamefont {C.}~\bibnamefont {Wang}}, \bibinfo {author} {\bibfnamefont {W.~R.}\ \bibnamefont {Sweeney}}, \bibinfo {author} {\bibfnamefont {A.~D.}\ \bibnamefont {Stone}},\ and\ \bibinfo {author} {\bibfnamefont {L.}~\bibnamefont {Yang}},\ }\bibfield  {title} {\bibinfo {title} {Coherent perfect absorption at an exceptional point},\ }\href {https://doi.org/10.1126/science.abj1028} {\bibfield  {journal} {\bibinfo  {journal} {Science}\ }\textbf {\bibinfo {volume} {373}},\ \bibinfo {pages} {1261} (\bibinfo {year} {2021})}\BibitemShut {NoStop}%
\bibitem [{\citenamefont {H{\"o}rner}\ \emph {et~al.}(2024)\citenamefont {H{\"o}rner}, \citenamefont {Wild}, \citenamefont {Slobodkin}, \citenamefont {Weinberg}, \citenamefont {Katz},\ and\ \citenamefont {Rotter}}]{horner2024Coherent}%
  \BibitemOpen
  \bibfield  {author} {\bibinfo {author} {\bibfnamefont {H.}~\bibnamefont {H{\"o}rner}}, \bibinfo {author} {\bibfnamefont {L.}~\bibnamefont {Wild}}, \bibinfo {author} {\bibfnamefont {Y.}~\bibnamefont {Slobodkin}}, \bibinfo {author} {\bibfnamefont {G.}~\bibnamefont {Weinberg}}, \bibinfo {author} {\bibfnamefont {O.}~\bibnamefont {Katz}},\ and\ \bibinfo {author} {\bibfnamefont {S.}~\bibnamefont {Rotter}},\ }\bibfield  {title} {\bibinfo {title} {Coherent {{Perfect Absorption}} of {{Arbitrary Wavefronts}} at an {{Exceptional Point}}},\ }\href {https://doi.org/10.1103/PhysRevLett.133.173801} {\bibfield  {journal} {\bibinfo  {journal} {Phys. Rev. Lett.}\ }\textbf {\bibinfo {volume} {133}},\ \bibinfo {pages} {173801} (\bibinfo {year} {2024})}\BibitemShut {NoStop}%
\bibitem [{\citenamefont {Xue}\ \emph {et~al.}(2025)\citenamefont {Xue}, \citenamefont {Lin}, \citenamefont {Hou}, \citenamefont {Zhu}, \citenamefont {Ma}, \citenamefont {Chen}, \citenamefont {Cheng}, \citenamefont {Ge},\ and\ \citenamefont {Wan}}]{xue2025DualColor}%
  \BibitemOpen
  \bibfield  {author} {\bibinfo {author} {\bibfnamefont {B.}~\bibnamefont {Xue}}, \bibinfo {author} {\bibfnamefont {J.}~\bibnamefont {Lin}}, \bibinfo {author} {\bibfnamefont {J.}~\bibnamefont {Hou}}, \bibinfo {author} {\bibfnamefont {Y.}~\bibnamefont {Zhu}}, \bibinfo {author} {\bibfnamefont {R.}~\bibnamefont {Ma}}, \bibinfo {author} {\bibfnamefont {X.}~\bibnamefont {Chen}}, \bibinfo {author} {\bibfnamefont {Y.}~\bibnamefont {Cheng}}, \bibinfo {author} {\bibfnamefont {L.}~\bibnamefont {Ge}},\ and\ \bibinfo {author} {\bibfnamefont {W.}~\bibnamefont {Wan}},\ }\bibfield  {title} {\bibinfo {title} {Dual-{{Color Coherent Perfect Absorber}}},\ }\href {https://doi.org/10.1103/PhysRevLett.134.013802} {\bibfield  {journal} {\bibinfo  {journal} {Phys. Rev. Lett.}\ }\textbf {\bibinfo {volume} {134}},\ \bibinfo {pages} {013802} (\bibinfo {year} {2025})}\BibitemShut {NoStop}%
\bibitem [{\citenamefont {Galiffi}\ \emph {et~al.}(2026)\citenamefont {Galiffi}, \citenamefont {Harwood}, \citenamefont {Vezzoli}, \citenamefont {Tirole}, \citenamefont {Al{\`u}},\ and\ \citenamefont {Sapienza}}]{galiffi2026Optical}%
  \BibitemOpen
  \bibfield  {author} {\bibinfo {author} {\bibfnamefont {E.}~\bibnamefont {Galiffi}}, \bibinfo {author} {\bibfnamefont {A.~C.}\ \bibnamefont {Harwood}}, \bibinfo {author} {\bibfnamefont {S.}~\bibnamefont {Vezzoli}}, \bibinfo {author} {\bibfnamefont {R.}~\bibnamefont {Tirole}}, \bibinfo {author} {\bibfnamefont {A.}~\bibnamefont {Al{\`u}}},\ and\ \bibinfo {author} {\bibfnamefont {R.}~\bibnamefont {Sapienza}},\ }\bibfield  {title} {\bibinfo {title} {Optical coherent perfect absorption and amplification in a time-varying medium},\ }\href {https://doi.org/10.1038/s41566-025-01833-8} {\bibfield  {journal} {\bibinfo  {journal} {Nat. Photonics}\ }\textbf {\bibinfo {volume} {20}},\ \bibinfo {pages} {163} (\bibinfo {year} {2026})}\BibitemShut {NoStop}%
\bibitem [{\citenamefont {Horodynski}\ \emph {et~al.}(2022)\citenamefont {Horodynski}, \citenamefont {K{\"u}hmayer}, \citenamefont {Ferise}, \citenamefont {Rotter},\ and\ \citenamefont {Davy}}]{horodynski2022Antireflection}%
  \BibitemOpen
  \bibfield  {author} {\bibinfo {author} {\bibfnamefont {M.}~\bibnamefont {Horodynski}}, \bibinfo {author} {\bibfnamefont {M.}~\bibnamefont {K{\"u}hmayer}}, \bibinfo {author} {\bibfnamefont {C.}~\bibnamefont {Ferise}}, \bibinfo {author} {\bibfnamefont {S.}~\bibnamefont {Rotter}},\ and\ \bibinfo {author} {\bibfnamefont {M.}~\bibnamefont {Davy}},\ }\bibfield  {title} {\bibinfo {title} {Anti-reflection structure for perfect transmission through complex media},\ }\href {https://doi.org/10.1038/s41586-022-04843-6} {\bibfield  {journal} {\bibinfo  {journal} {Nature}\ }\textbf {\bibinfo {volume} {607}},\ \bibinfo {pages} {281} (\bibinfo {year} {2022})}\BibitemShut {NoStop}%
\bibitem [{\citenamefont {Noh}(2012)}]{noh2012Perfecta}%
  \BibitemOpen
  \bibfield  {author} {\bibinfo {author} {\bibfnamefont {H.}~\bibnamefont {Noh}},\ }\bibfield  {title} {\bibinfo {title} {Perfect coupling of light to surface plasmons by coherent absorption},\ }\href {https://doi.org/10.1103/PhysRevLett.108.186805} {\bibfield  {journal} {\bibinfo  {journal} {Phys. Rev. Lett.}\ }\textbf {\bibinfo {volume} {108}},\ \bibinfo {pages} {186805} (\bibinfo {year} {2012})}\BibitemShut {NoStop}%
\bibitem [{\citenamefont {Roger}\ \emph {et~al.}(2015)\citenamefont {Roger}, \citenamefont {Vezzoli}, \citenamefont {Bolduc}, \citenamefont {Valente}, \citenamefont {Heitz}, \citenamefont {Jeffers}, \citenamefont {Soci}, \citenamefont {Leach}, \citenamefont {Couteau}, \citenamefont {Zheludev},\ and\ \citenamefont {Faccio}}]{roger2015Coherent}%
  \BibitemOpen
  \bibfield  {author} {\bibinfo {author} {\bibfnamefont {T.}~\bibnamefont {Roger}}, \bibinfo {author} {\bibfnamefont {S.}~\bibnamefont {Vezzoli}}, \bibinfo {author} {\bibfnamefont {E.}~\bibnamefont {Bolduc}}, \bibinfo {author} {\bibfnamefont {J.}~\bibnamefont {Valente}}, \bibinfo {author} {\bibfnamefont {J.~J.~F.}\ \bibnamefont {Heitz}}, \bibinfo {author} {\bibfnamefont {J.}~\bibnamefont {Jeffers}}, \bibinfo {author} {\bibfnamefont {C.}~\bibnamefont {Soci}}, \bibinfo {author} {\bibfnamefont {J.}~\bibnamefont {Leach}}, \bibinfo {author} {\bibfnamefont {C.}~\bibnamefont {Couteau}}, \bibinfo {author} {\bibfnamefont {N.~I.}\ \bibnamefont {Zheludev}},\ and\ \bibinfo {author} {\bibfnamefont {D.}~\bibnamefont {Faccio}},\ }\bibfield  {title} {\bibinfo {title} {Coherent perfect absorption in deeply subwavelength films in the single-photon regime},\ }\href {https://doi.org/10.1038/ncomms8031} {\bibfield  {journal} {\bibinfo  {journal} {Nat. Commun.}\ }\textbf {\bibinfo {volume} {6}},\ \bibinfo {pages} {7031} (\bibinfo
  {year} {2015})}\BibitemShut {NoStop}%
\bibitem [{\citenamefont {Ma}\ \emph {et~al.}(2014)\citenamefont {Ma}, \citenamefont {Yang}, \citenamefont {Xiao}, \citenamefont {Yang},\ and\ \citenamefont {Sheng}}]{ma2014Acoustic}%
  \BibitemOpen
  \bibfield  {author} {\bibinfo {author} {\bibfnamefont {G.}~\bibnamefont {Ma}}, \bibinfo {author} {\bibfnamefont {M.}~\bibnamefont {Yang}}, \bibinfo {author} {\bibfnamefont {S.}~\bibnamefont {Xiao}}, \bibinfo {author} {\bibfnamefont {Z.}~\bibnamefont {Yang}},\ and\ \bibinfo {author} {\bibfnamefont {P.}~\bibnamefont {Sheng}},\ }\bibfield  {title} {\bibinfo {title} {Acoustic metasurface with hybrid resonances},\ }\href {https://doi.org/10.1038/nmat3994} {\bibfield  {journal} {\bibinfo  {journal} {Nat. Mater.}\ }\textbf {\bibinfo {volume} {13}},\ \bibinfo {pages} {873} (\bibinfo {year} {2014})}\BibitemShut {NoStop}%
\bibitem [{\citenamefont {Xia}\ \emph {et~al.}(2025)\citenamefont {Xia}, \citenamefont {Xu}, \citenamefont {Yan}, \citenamefont {Chen}, \citenamefont {Yang}, \citenamefont {Liang}, \citenamefont {Cheng},\ and\ \citenamefont {Christensen}}]{xia2025Observation}%
  \BibitemOpen
  \bibfield  {author} {\bibinfo {author} {\bibfnamefont {Y.-F.}\ \bibnamefont {Xia}}, \bibinfo {author} {\bibfnamefont {Z.-X.}\ \bibnamefont {Xu}}, \bibinfo {author} {\bibfnamefont {Y.-T.}\ \bibnamefont {Yan}}, \bibinfo {author} {\bibfnamefont {A.}~\bibnamefont {Chen}}, \bibinfo {author} {\bibfnamefont {J.}~\bibnamefont {Yang}}, \bibinfo {author} {\bibfnamefont {B.}~\bibnamefont {Liang}}, \bibinfo {author} {\bibfnamefont {J.-C.}\ \bibnamefont {Cheng}},\ and\ \bibinfo {author} {\bibfnamefont {J.}~\bibnamefont {Christensen}},\ }\bibfield  {title} {\bibinfo {title} {Observation of {{Coherent Perfect Acoustic Absorption}} at an {{Exceptional Point}}},\ }\href {https://doi.org/10.1103/slhy-f76q} {\bibfield  {journal} {\bibinfo  {journal} {Phys. Rev. Lett.}\ }\textbf {\bibinfo {volume} {135}},\ \bibinfo {pages} {067001} (\bibinfo {year} {2025})}\BibitemShut {NoStop}%
\bibitem [{\citenamefont {Sun}\ \emph {et~al.}(2014)\citenamefont {Sun}, \citenamefont {Tan}, \citenamefont {Li}, \citenamefont {Li},\ and\ \citenamefont {Chen}}]{sun2014Experimentala}%
  \BibitemOpen
  \bibfield  {author} {\bibinfo {author} {\bibfnamefont {Y.}~\bibnamefont {Sun}}, \bibinfo {author} {\bibfnamefont {W.}~\bibnamefont {Tan}}, \bibinfo {author} {\bibfnamefont {H.-q.}\ \bibnamefont {Li}}, \bibinfo {author} {\bibfnamefont {J.}~\bibnamefont {Li}},\ and\ \bibinfo {author} {\bibfnamefont {H.}~\bibnamefont {Chen}},\ }\bibfield  {title} {\bibinfo {title} {Experimental demonstration of a coherent perfect absorber with pt phase transition},\ }\href {https://doi.org/10.1103/PhysRevLett.112.143903} {\bibfield  {journal} {\bibinfo  {journal} {Phys. Rev. Lett.}\ }\textbf {\bibinfo {volume} {112}},\ \bibinfo {pages} {143903} (\bibinfo {year} {2014})}\BibitemShut {NoStop}%
\bibitem [{\citenamefont {Feng}\ \emph {et~al.}(2017)\citenamefont {Feng}, \citenamefont {{El-Ganainy}},\ and\ \citenamefont {Ge}}]{feng2017NonHermitian}%
  \BibitemOpen
  \bibfield  {author} {\bibinfo {author} {\bibfnamefont {L.}~\bibnamefont {Feng}}, \bibinfo {author} {\bibfnamefont {R.}~\bibnamefont {{El-Ganainy}}},\ and\ \bibinfo {author} {\bibfnamefont {L.}~\bibnamefont {Ge}},\ }\bibfield  {title} {\bibinfo {title} {Non-{{Hermitian}} photonics based on parity--time symmetry},\ }\href {https://doi.org/10.1038/s41566-017-0031-1} {\bibfield  {journal} {\bibinfo  {journal} {Nat. Photonics}\ }\textbf {\bibinfo {volume} {11}},\ \bibinfo {pages} {752} (\bibinfo {year} {2017})}\BibitemShut {NoStop}%
\bibitem [{\citenamefont {Noh}\ \emph {et~al.}(2012)\citenamefont {Noh}, \citenamefont {Chong}, \citenamefont {Stone},\ and\ \citenamefont {Cao}}]{noh2012Perfect}%
  \BibitemOpen
  \bibfield  {author} {\bibinfo {author} {\bibfnamefont {H.}~\bibnamefont {Noh}}, \bibinfo {author} {\bibfnamefont {Y.}~\bibnamefont {Chong}}, \bibinfo {author} {\bibfnamefont {A.~D.}\ \bibnamefont {Stone}},\ and\ \bibinfo {author} {\bibfnamefont {H.}~\bibnamefont {Cao}},\ }\bibfield  {title} {\bibinfo {title} {Perfect coupling of light to surface plasmons by coherent absorption},\ }\href {https://doi.org/10.1103/PhysRevLett.108.186805} {\bibfield  {journal} {\bibinfo  {journal} {Phys. Rev. Lett.}\ }\textbf {\bibinfo {volume} {108}},\ \bibinfo {pages} {186805} (\bibinfo {year} {2012})}\BibitemShut {NoStop}%
\bibitem [{\citenamefont {Zanotto}\ \emph {et~al.}(2014)\citenamefont {Zanotto}, \citenamefont {Mezzapesa}, \citenamefont {Bianco}, \citenamefont {Biasiol}, \citenamefont {Baldacci}, \citenamefont {Vitiello}, \citenamefont {Sorba}, \citenamefont {Colombelli},\ and\ \citenamefont {Tredicucci}}]{zanotto2014Perfect}%
  \BibitemOpen
  \bibfield  {author} {\bibinfo {author} {\bibfnamefont {S.}~\bibnamefont {Zanotto}}, \bibinfo {author} {\bibfnamefont {F.~P.}\ \bibnamefont {Mezzapesa}}, \bibinfo {author} {\bibfnamefont {F.}~\bibnamefont {Bianco}}, \bibinfo {author} {\bibfnamefont {G.}~\bibnamefont {Biasiol}}, \bibinfo {author} {\bibfnamefont {L.}~\bibnamefont {Baldacci}}, \bibinfo {author} {\bibfnamefont {M.~S.}\ \bibnamefont {Vitiello}}, \bibinfo {author} {\bibfnamefont {L.}~\bibnamefont {Sorba}}, \bibinfo {author} {\bibfnamefont {R.}~\bibnamefont {Colombelli}},\ and\ \bibinfo {author} {\bibfnamefont {A.}~\bibnamefont {Tredicucci}},\ }\bibfield  {title} {\bibinfo {title} {Perfect energy-feeding into strongly coupled systems and interferometric control of polariton absorption},\ }\href {https://doi.org/10.1038/nphys3106} {\bibfield  {journal} {\bibinfo  {journal} {Nat. Phys.}\ }\textbf {\bibinfo {volume} {10}},\ \bibinfo {pages} {830} (\bibinfo {year} {2014})}\BibitemShut {NoStop}%
\bibitem [{\citenamefont {Zhang}\ \emph {et~al.}(2017)\citenamefont {Zhang}, \citenamefont {Luo}, \citenamefont {Wang}, \citenamefont {Li},\ and\ \citenamefont {You}}]{zhang2017Observationa}%
  \BibitemOpen
  \bibfield  {author} {\bibinfo {author} {\bibfnamefont {D.}~\bibnamefont {Zhang}}, \bibinfo {author} {\bibfnamefont {X.-Q.}\ \bibnamefont {Luo}}, \bibinfo {author} {\bibfnamefont {Y.-P.}\ \bibnamefont {Wang}}, \bibinfo {author} {\bibfnamefont {T.-F.}\ \bibnamefont {Li}},\ and\ \bibinfo {author} {\bibfnamefont {J.~Q.}\ \bibnamefont {You}},\ }\bibfield  {title} {\bibinfo {title} {Observation of the exceptional point in cavity magnon-polaritons},\ }\href {https://doi.org/10.1038/s41467-017-01634-w} {\bibfield  {journal} {\bibinfo  {journal} {Nat. Commun.}\ }\textbf {\bibinfo {volume} {8}},\ \bibinfo {pages} {1368} (\bibinfo {year} {2017})}\BibitemShut {NoStop}%
\bibitem [{\citenamefont {Weis}\ \emph {et~al.}(2010)\citenamefont {Weis}, \citenamefont {Rivi{\`e}re}, \citenamefont {Del{\'e}glise}, \citenamefont {Gavartin}, \citenamefont {Arcizet}, \citenamefont {Schliesser},\ and\ \citenamefont {Kippenberg}}]{weis2010Optomechanically}%
  \BibitemOpen
  \bibfield  {author} {\bibinfo {author} {\bibfnamefont {S.}~\bibnamefont {Weis}}, \bibinfo {author} {\bibfnamefont {R.}~\bibnamefont {Rivi{\`e}re}}, \bibinfo {author} {\bibfnamefont {S.}~\bibnamefont {Del{\'e}glise}}, \bibinfo {author} {\bibfnamefont {E.}~\bibnamefont {Gavartin}}, \bibinfo {author} {\bibfnamefont {O.}~\bibnamefont {Arcizet}}, \bibinfo {author} {\bibfnamefont {A.}~\bibnamefont {Schliesser}},\ and\ \bibinfo {author} {\bibfnamefont {T.~J.}\ \bibnamefont {Kippenberg}},\ }\bibfield  {title} {\bibinfo {title} {Optomechanically {{Induced Transparency}}},\ }\href {https://doi.org/10.1126/science.1195596} {\bibfield  {journal} {\bibinfo  {journal} {Science}\ }\textbf {\bibinfo {volume} {330}},\ \bibinfo {pages} {1520} (\bibinfo {year} {2010})}\BibitemShut {NoStop}%
\bibitem [{\citenamefont {Massel}\ \emph {et~al.}(2011)\citenamefont {Massel}, \citenamefont {Heikkil{\"a}}, \citenamefont {Pirkkalainen}, \citenamefont {Cho}, \citenamefont {Saloniemi}, \citenamefont {Hakonen},\ and\ \citenamefont {Sillanp{\"a}{\"a}}}]{massel2011Microwave}%
  \BibitemOpen
  \bibfield  {author} {\bibinfo {author} {\bibfnamefont {F.}~\bibnamefont {Massel}}, \bibinfo {author} {\bibfnamefont {T.~T.}\ \bibnamefont {Heikkil{\"a}}}, \bibinfo {author} {\bibfnamefont {J.-M.}\ \bibnamefont {Pirkkalainen}}, \bibinfo {author} {\bibfnamefont {S.~U.}\ \bibnamefont {Cho}}, \bibinfo {author} {\bibfnamefont {H.}~\bibnamefont {Saloniemi}}, \bibinfo {author} {\bibfnamefont {P.~J.}\ \bibnamefont {Hakonen}},\ and\ \bibinfo {author} {\bibfnamefont {M.~A.}\ \bibnamefont {Sillanp{\"a}{\"a}}},\ }\bibfield  {title} {\bibinfo {title} {Microwave amplification with nanomechanical resonators},\ }\href {https://doi.org/10.1038/nature10628} {\bibfield  {journal} {\bibinfo  {journal} {Nature}\ }\textbf {\bibinfo {volume} {480}},\ \bibinfo {pages} {351} (\bibinfo {year} {2011})}\BibitemShut {NoStop}%
\bibitem [{\citenamefont {{Safavi-Naeini}}\ \emph {et~al.}(2011)\citenamefont {{Safavi-Naeini}}, \citenamefont {Alegre}, \citenamefont {Chan}, \citenamefont {Eichenfield}, \citenamefont {Winger}, \citenamefont {Lin}, \citenamefont {Hill}, \citenamefont {Chang},\ and\ \citenamefont {Painter}}]{safavi-naeini2011Electromagnetically}%
  \BibitemOpen
  \bibfield  {author} {\bibinfo {author} {\bibfnamefont {A.~H.}\ \bibnamefont {{Safavi-Naeini}}}, \bibinfo {author} {\bibfnamefont {T.~P.~M.}\ \bibnamefont {Alegre}}, \bibinfo {author} {\bibfnamefont {J.}~\bibnamefont {Chan}}, \bibinfo {author} {\bibfnamefont {M.}~\bibnamefont {Eichenfield}}, \bibinfo {author} {\bibfnamefont {M.}~\bibnamefont {Winger}}, \bibinfo {author} {\bibfnamefont {Q.}~\bibnamefont {Lin}}, \bibinfo {author} {\bibfnamefont {J.~T.}\ \bibnamefont {Hill}}, \bibinfo {author} {\bibfnamefont {D.~E.}\ \bibnamefont {Chang}},\ and\ \bibinfo {author} {\bibfnamefont {O.}~\bibnamefont {Painter}},\ }\bibfield  {title} {\bibinfo {title} {Electromagnetically induced transparency and slow light with optomechanics},\ }\href {https://doi.org/10.1038/nature09933} {\bibfield  {journal} {\bibinfo  {journal} {Nature}\ }\textbf {\bibinfo {volume} {472}},\ \bibinfo {pages} {69} (\bibinfo {year} {2011})}\BibitemShut {NoStop}%
\bibitem [{\citenamefont {Hocke}\ \emph {et~al.}(2012)\citenamefont {Hocke}, \citenamefont {Zhou}, \citenamefont {Schliesser}, \citenamefont {Kippenberg}, \citenamefont {Huebl},\ and\ \citenamefont {Gross}}]{hocke2012Electromechanically}%
  \BibitemOpen
  \bibfield  {author} {\bibinfo {author} {\bibfnamefont {F.}~\bibnamefont {Hocke}}, \bibinfo {author} {\bibfnamefont {X.}~\bibnamefont {Zhou}}, \bibinfo {author} {\bibfnamefont {A.}~\bibnamefont {Schliesser}}, \bibinfo {author} {\bibfnamefont {T.~J.}\ \bibnamefont {Kippenberg}}, \bibinfo {author} {\bibfnamefont {H.}~\bibnamefont {Huebl}},\ and\ \bibinfo {author} {\bibfnamefont {R.}~\bibnamefont {Gross}},\ }\bibfield  {title} {\bibinfo {title} {Electromechanically induced absorption in a circuit nano-electromechanical system},\ }\href {https://doi.org/10.1088/1367-2630/14/12/123037} {\bibfield  {journal} {\bibinfo  {journal} {New J. Phys.}\ }\textbf {\bibinfo {volume} {14}},\ \bibinfo {pages} {123037} (\bibinfo {year} {2012})}\BibitemShut {NoStop}%
\bibitem [{\citenamefont {Zhou}\ \emph {et~al.}(2013)\citenamefont {Zhou}, \citenamefont {Hocke}, \citenamefont {Schliesser}, \citenamefont {Marx}, \citenamefont {Huebl}, \citenamefont {Gross},\ and\ \citenamefont {Kippenberg}}]{zhou2013Slowing}%
  \BibitemOpen
  \bibfield  {author} {\bibinfo {author} {\bibfnamefont {X.}~\bibnamefont {Zhou}}, \bibinfo {author} {\bibfnamefont {F.}~\bibnamefont {Hocke}}, \bibinfo {author} {\bibfnamefont {A.}~\bibnamefont {Schliesser}}, \bibinfo {author} {\bibfnamefont {A.}~\bibnamefont {Marx}}, \bibinfo {author} {\bibfnamefont {H.}~\bibnamefont {Huebl}}, \bibinfo {author} {\bibfnamefont {R.}~\bibnamefont {Gross}},\ and\ \bibinfo {author} {\bibfnamefont {T.~J.}\ \bibnamefont {Kippenberg}},\ }\bibfield  {title} {\bibinfo {title} {Slowing, advancing and switching of microwave signals using circuit nanoelectromechanics},\ }\href {https://doi.org/10.1038/nphys2527} {\bibfield  {journal} {\bibinfo  {journal} {Nat. Phys.}\ }\textbf {\bibinfo {volume} {9}},\ \bibinfo {pages} {179} (\bibinfo {year} {2013})}\BibitemShut {NoStop}%
\bibitem [{\citenamefont {Pendry}\ \emph {et~al.}(2006)\citenamefont {Pendry}, \citenamefont {Schurig},\ and\ \citenamefont {Smith}}]{pendry2006Controlling}%
  \BibitemOpen
  \bibfield  {author} {\bibinfo {author} {\bibfnamefont {J.~B.}\ \bibnamefont {Pendry}}, \bibinfo {author} {\bibfnamefont {D.}~\bibnamefont {Schurig}},\ and\ \bibinfo {author} {\bibfnamefont {D.~R.}\ \bibnamefont {Smith}},\ }\bibfield  {title} {\bibinfo {title} {Controlling {{Electromagnetic Fields}}},\ }\href {https://doi.org/10.1126/science.1125907} {\bibfield  {journal} {\bibinfo  {journal} {Science}\ }\textbf {\bibinfo {volume} {312}},\ \bibinfo {pages} {1780} (\bibinfo {year} {2006})}\BibitemShut {NoStop}%
\bibitem [{\citenamefont {Rechtsman}(2017)}]{rechtsman2017Optical}%
  \BibitemOpen
  \bibfield  {author} {\bibinfo {author} {\bibfnamefont {M.~C.}\ \bibnamefont {Rechtsman}},\ }\bibfield  {title} {\bibinfo {title} {Optical sensing gets exceptional},\ }\href {https://doi.org/10.1038/548161a} {\bibfield  {journal} {\bibinfo  {journal} {Nature}\ }\textbf {\bibinfo {volume} {548}},\ \bibinfo {pages} {161} (\bibinfo {year} {2017})}\BibitemShut {NoStop}%
\bibitem [{\citenamefont {Liu}\ \emph {et~al.}(2010)\citenamefont {Liu}, \citenamefont {Mesch}, \citenamefont {Weiss}, \citenamefont {Hentschel},\ and\ \citenamefont {Giessen}}]{liu2010Infrared}%
  \BibitemOpen
  \bibfield  {author} {\bibinfo {author} {\bibfnamefont {N.}~\bibnamefont {Liu}}, \bibinfo {author} {\bibfnamefont {M.}~\bibnamefont {Mesch}}, \bibinfo {author} {\bibfnamefont {T.}~\bibnamefont {Weiss}}, \bibinfo {author} {\bibfnamefont {M.}~\bibnamefont {Hentschel}},\ and\ \bibinfo {author} {\bibfnamefont {H.}~\bibnamefont {Giessen}},\ }\bibfield  {title} {\bibinfo {title} {Infrared {{Perfect Absorber}} and {{Its Application As Plasmonic Sensor}}},\ }\href {https://doi.org/10.1021/nl9041033} {\bibfield  {journal} {\bibinfo  {journal} {Nano Lett.}\ }\textbf {\bibinfo {volume} {10}},\ \bibinfo {pages} {2342} (\bibinfo {year} {2010})}\BibitemShut {NoStop}%
\bibitem [{\citenamefont {Vetlugin}(2021)}]{vetlugin2021Coherent}%
  \BibitemOpen
  \bibfield  {author} {\bibinfo {author} {\bibfnamefont {A.~N.}\ \bibnamefont {Vetlugin}},\ }\bibfield  {title} {\bibinfo {title} {Coherent perfect absorption of quantum light},\ }\href {https://doi.org/10.1103/PhysRevA.104.013716} {\bibfield  {journal} {\bibinfo  {journal} {Phys. Rev. A}\ }\textbf {\bibinfo {volume} {104}},\ \bibinfo {pages} {013716} (\bibinfo {year} {2021})}\BibitemShut {NoStop}%
\bibitem [{\citenamefont {Vest}\ \emph {et~al.}(2017)\citenamefont {Vest}, \citenamefont {Dheur}, \citenamefont {Devaux}, \citenamefont {Baron}, \citenamefont {Rousseau}, \citenamefont {Hugonin}, \citenamefont {Greffet}, \citenamefont {Messin},\ and\ \citenamefont {Marquier}}]{vest2017Anticoalescence}%
  \BibitemOpen
  \bibfield  {author} {\bibinfo {author} {\bibfnamefont {B.}~\bibnamefont {Vest}}, \bibinfo {author} {\bibfnamefont {M.-C.}\ \bibnamefont {Dheur}}, \bibinfo {author} {\bibfnamefont {{\'E}.}~\bibnamefont {Devaux}}, \bibinfo {author} {\bibfnamefont {A.}~\bibnamefont {Baron}}, \bibinfo {author} {\bibfnamefont {E.}~\bibnamefont {Rousseau}}, \bibinfo {author} {\bibfnamefont {J.-P.}\ \bibnamefont {Hugonin}}, \bibinfo {author} {\bibfnamefont {J.-J.}\ \bibnamefont {Greffet}}, \bibinfo {author} {\bibfnamefont {G.}~\bibnamefont {Messin}},\ and\ \bibinfo {author} {\bibfnamefont {F.}~\bibnamefont {Marquier}},\ }\bibfield  {title} {\bibinfo {title} {Anti-coalescence of bosons on a lossy beam splitter},\ }\href {https://doi.org/10.1126/science.aam9353} {\bibfield  {journal} {\bibinfo  {journal} {Science}\ }\textbf {\bibinfo {volume} {356}},\ \bibinfo {pages} {1373} (\bibinfo {year} {2017})}\BibitemShut {NoStop}%
\bibitem [{\citenamefont {Roger}\ \emph {et~al.}(2016)\citenamefont {Roger}, \citenamefont {Restuccia}, \citenamefont {Lyons}, \citenamefont {Giovannini}, \citenamefont {Romero}, \citenamefont {Jeffers}, \citenamefont {Padgett},\ and\ \citenamefont {Faccio}}]{roger2016Coherent}%
  \BibitemOpen
  \bibfield  {author} {\bibinfo {author} {\bibfnamefont {T.}~\bibnamefont {Roger}}, \bibinfo {author} {\bibfnamefont {S.}~\bibnamefont {Restuccia}}, \bibinfo {author} {\bibfnamefont {A.}~\bibnamefont {Lyons}}, \bibinfo {author} {\bibfnamefont {D.}~\bibnamefont {Giovannini}}, \bibinfo {author} {\bibfnamefont {J.}~\bibnamefont {Romero}}, \bibinfo {author} {\bibfnamefont {J.}~\bibnamefont {Jeffers}}, \bibinfo {author} {\bibfnamefont {M.}~\bibnamefont {Padgett}},\ and\ \bibinfo {author} {\bibfnamefont {D.}~\bibnamefont {Faccio}},\ }\bibfield  {title} {\bibinfo {title} {Coherent absorption of {N00N} states},\ }\href {https://doi.org/10.1103/PhysRevLett.117.023601} {\bibfield  {journal} {\bibinfo  {journal} {Phys. Rev. Lett.}\ }\textbf {\bibinfo {volume} {117}},\ \bibinfo {pages} {023601} (\bibinfo {year} {2016})}\BibitemShut {NoStop}%
\bibitem [{\citenamefont {Altuzarra}\ \emph {et~al.}(2017)\citenamefont {Altuzarra}, \citenamefont {Vezzoli}, \citenamefont {Valente}, \citenamefont {Gao}, \citenamefont {Soci}, \citenamefont {Faccio},\ and\ \citenamefont {Couteau}}]{altuzarra2017Coherent}%
  \BibitemOpen
  \bibfield  {author} {\bibinfo {author} {\bibfnamefont {C.}~\bibnamefont {Altuzarra}}, \bibinfo {author} {\bibfnamefont {S.}~\bibnamefont {Vezzoli}}, \bibinfo {author} {\bibfnamefont {J.}~\bibnamefont {Valente}}, \bibinfo {author} {\bibfnamefont {W.}~\bibnamefont {Gao}}, \bibinfo {author} {\bibfnamefont {C.}~\bibnamefont {Soci}}, \bibinfo {author} {\bibfnamefont {D.}~\bibnamefont {Faccio}},\ and\ \bibinfo {author} {\bibfnamefont {C.}~\bibnamefont {Couteau}},\ }\bibfield  {title} {\bibinfo {title} {Coherent perfect absorption in metamaterials with entangled photons},\ }\href {https://doi.org/10.1021/acsphotonics.7b00514} {\bibfield  {journal} {\bibinfo  {journal} {ACS Photonics}\ }\textbf {\bibinfo {volume} {4}},\ \bibinfo {pages} {2124} (\bibinfo {year} {2017})}\BibitemShut {NoStop}%
\bibitem [{\citenamefont {Hern{\'a}ndez}\ \emph {et~al.}(2022)\citenamefont {Hern{\'a}ndez}, \citenamefont {{Ortega-Gomez}}, \citenamefont {Bravo},\ and\ \citenamefont {Liberal}}]{hernandez2022Quantum}%
  \BibitemOpen
  \bibfield  {author} {\bibinfo {author} {\bibfnamefont {O.}~\bibnamefont {Hern{\'a}ndez}}, \bibinfo {author} {\bibfnamefont {A.}~\bibnamefont {{Ortega-Gomez}}}, \bibinfo {author} {\bibfnamefont {M.}~\bibnamefont {Bravo}},\ and\ \bibinfo {author} {\bibfnamefont {I.}~\bibnamefont {Liberal}},\ }\bibfield  {title} {\bibinfo {title} {Quantum interference in wilkinson power dividers},\ }\href {https://doi.org/10.1002/lpor.202200095} {\bibfield  {journal} {\bibinfo  {journal} {Laser Photonics Rev.}\ }\textbf {\bibinfo {volume} {16}},\ \bibinfo {pages} {2200095} (\bibinfo {year} {2022})}\BibitemShut {NoStop}%
\bibitem [{\citenamefont {M{\"u}llers}\ \emph {et~al.}(2018)\citenamefont {M{\"u}llers}, \citenamefont {Santra}, \citenamefont {Baals}, \citenamefont {Jiang}, \citenamefont {Benary}, \citenamefont {Labouvie}, \citenamefont {Zezyulin}, \citenamefont {Konotop},\ and\ \citenamefont {Ott}}]{mullers2018Coherent}%
  \BibitemOpen
  \bibfield  {author} {\bibinfo {author} {\bibfnamefont {A.}~\bibnamefont {M{\"u}llers}}, \bibinfo {author} {\bibfnamefont {B.}~\bibnamefont {Santra}}, \bibinfo {author} {\bibfnamefont {C.}~\bibnamefont {Baals}}, \bibinfo {author} {\bibfnamefont {J.}~\bibnamefont {Jiang}}, \bibinfo {author} {\bibfnamefont {J.}~\bibnamefont {Benary}}, \bibinfo {author} {\bibfnamefont {R.}~\bibnamefont {Labouvie}}, \bibinfo {author} {\bibfnamefont {D.~A.}\ \bibnamefont {Zezyulin}}, \bibinfo {author} {\bibfnamefont {V.~V.}\ \bibnamefont {Konotop}},\ and\ \bibinfo {author} {\bibfnamefont {H.}~\bibnamefont {Ott}},\ }\bibfield  {title} {\bibinfo {title} {Coherent perfect absorption of nonlinear matter waves},\ }\href {https://doi.org/10.1126/sciadv.aat6539} {\bibfield  {journal} {\bibinfo  {journal} {Sci. Adv.}\ }\textbf {\bibinfo {volume} {4}},\ \bibinfo {pages} {eaat6539} (\bibinfo {year} {2018})}\BibitemShut {NoStop}%
\bibitem [{\citenamefont {Lai}\ \emph {et~al.}(2024)\citenamefont {Lai}, \citenamefont {Clarke}, \citenamefont {Grimm}, \citenamefont {Devi}, \citenamefont {Wigger}, \citenamefont {Helbig}, \citenamefont {Hofmann}, \citenamefont {Thomale}, \citenamefont {Huang}, \citenamefont {Hecht},\ and\ \citenamefont {Hess}}]{lai2024Roomtemperature}%
  \BibitemOpen
  \bibfield  {author} {\bibinfo {author} {\bibfnamefont {Y.}~\bibnamefont {Lai}}, \bibinfo {author} {\bibfnamefont {D.~D.~A.}\ \bibnamefont {Clarke}}, \bibinfo {author} {\bibfnamefont {P.}~\bibnamefont {Grimm}}, \bibinfo {author} {\bibfnamefont {A.}~\bibnamefont {Devi}}, \bibinfo {author} {\bibfnamefont {D.}~\bibnamefont {Wigger}}, \bibinfo {author} {\bibfnamefont {T.}~\bibnamefont {Helbig}}, \bibinfo {author} {\bibfnamefont {T.}~\bibnamefont {Hofmann}}, \bibinfo {author} {\bibfnamefont {R.}~\bibnamefont {Thomale}}, \bibinfo {author} {\bibfnamefont {J.-S.}\ \bibnamefont {Huang}}, \bibinfo {author} {\bibfnamefont {B.}~\bibnamefont {Hecht}},\ and\ \bibinfo {author} {\bibfnamefont {O.}~\bibnamefont {Hess}},\ }\bibfield  {title} {\bibinfo {title} {Room-temperature quantum nanoplasmonic coherent perfect absorption},\ }\href {https://doi.org/10.1038/s41467-024-50574-9} {\bibfield  {journal} {\bibinfo  {journal} {Nat. Commun.}\ }\textbf {\bibinfo {volume} {15}},\ \bibinfo {pages} {6324} (\bibinfo {year}
  {2024})}\BibitemShut {NoStop}%
\bibitem [{\citenamefont {Aspelmeyer}\ \emph {et~al.}(2014)\citenamefont {Aspelmeyer}, \citenamefont {Kippenberg},\ and\ \citenamefont {Marquardt}}]{aspelmeyer2014Cavityb}%
  \BibitemOpen
  \bibfield  {author} {\bibinfo {author} {\bibfnamefont {M.}~\bibnamefont {Aspelmeyer}}, \bibinfo {author} {\bibfnamefont {T.~J.}\ \bibnamefont {Kippenberg}},\ and\ \bibinfo {author} {\bibfnamefont {F.}~\bibnamefont {Marquardt}},\ }\bibfield  {title} {\bibinfo {title} {Cavity optomechanics},\ }\href {https://doi.org/10.1103/RevModPhys.86.1391} {\bibfield  {journal} {\bibinfo  {journal} {Rev. Mod. Phys.}\ }\textbf {\bibinfo {volume} {86}},\ \bibinfo {pages} {1391} (\bibinfo {year} {2014})}\BibitemShut {NoStop}%
\bibitem [{\citenamefont {Pirkkalainen}\ \emph {et~al.}(2013)\citenamefont {Pirkkalainen}, \citenamefont {Cho}, \citenamefont {Li}, \citenamefont {Paraoanu}, \citenamefont {Hakonen},\ and\ \citenamefont {Sillanp{\"a}{\"a}}}]{pirkkalainen2013Hybrid}%
  \BibitemOpen
  \bibfield  {author} {\bibinfo {author} {\bibfnamefont {J.-M.}\ \bibnamefont {Pirkkalainen}}, \bibinfo {author} {\bibfnamefont {S.~U.}\ \bibnamefont {Cho}}, \bibinfo {author} {\bibfnamefont {J.}~\bibnamefont {Li}}, \bibinfo {author} {\bibfnamefont {G.~S.}\ \bibnamefont {Paraoanu}}, \bibinfo {author} {\bibfnamefont {P.~J.}\ \bibnamefont {Hakonen}},\ and\ \bibinfo {author} {\bibfnamefont {M.~A.}\ \bibnamefont {Sillanp{\"a}{\"a}}},\ }\bibfield  {title} {\bibinfo {title} {Hybrid circuit cavity quantum electrodynamics with a micromechanical resonator},\ }\href {https://doi.org/10.1038/nature11821} {\bibfield  {journal} {\bibinfo  {journal} {Nature}\ }\textbf {\bibinfo {volume} {494}},\ \bibinfo {pages} {211} (\bibinfo {year} {2013})}\BibitemShut {NoStop}%
\bibitem [{\citenamefont {O'Connell}\ \emph {et~al.}(2010)\citenamefont {O'Connell}, \citenamefont {Hofheinz}, \citenamefont {Ansmann}, \citenamefont {Bialczak}, \citenamefont {Lenander}, \citenamefont {Lucero}, \citenamefont {Neeley}, \citenamefont {Sank}, \citenamefont {Wang}, \citenamefont {Weides}, \citenamefont {Wenner}, \citenamefont {Martinis},\ and\ \citenamefont {Cleland}}]{oconnell2010Quantum}%
  \BibitemOpen
  \bibfield  {author} {\bibinfo {author} {\bibfnamefont {A.~D.}\ \bibnamefont {O'Connell}}, \bibinfo {author} {\bibfnamefont {M.}~\bibnamefont {Hofheinz}}, \bibinfo {author} {\bibfnamefont {M.}~\bibnamefont {Ansmann}}, \bibinfo {author} {\bibfnamefont {R.~C.}\ \bibnamefont {Bialczak}}, \bibinfo {author} {\bibfnamefont {M.}~\bibnamefont {Lenander}}, \bibinfo {author} {\bibfnamefont {E.}~\bibnamefont {Lucero}}, \bibinfo {author} {\bibfnamefont {M.}~\bibnamefont {Neeley}}, \bibinfo {author} {\bibfnamefont {D.}~\bibnamefont {Sank}}, \bibinfo {author} {\bibfnamefont {H.}~\bibnamefont {Wang}}, \bibinfo {author} {\bibfnamefont {M.}~\bibnamefont {Weides}}, \bibinfo {author} {\bibfnamefont {J.}~\bibnamefont {Wenner}}, \bibinfo {author} {\bibfnamefont {J.~M.}\ \bibnamefont {Martinis}},\ and\ \bibinfo {author} {\bibfnamefont {A.~N.}\ \bibnamefont {Cleland}},\ }\bibfield  {title} {\bibinfo {title} {Quantum ground state and single-phonon control of a mechanical resonator},\ }\href {https://doi.org/10.1038/nature08967}
  {\bibfield  {journal} {\bibinfo  {journal} {Nature}\ }\textbf {\bibinfo {volume} {464}},\ \bibinfo {pages} {697} (\bibinfo {year} {2010})}\BibitemShut {NoStop}%
\bibitem [{\citenamefont {Liu}\ \emph {et~al.}(2021)\citenamefont {Liu}, \citenamefont {Liu}, \citenamefont {Wang}, \citenamefont {Chen}, \citenamefont {Sillanp{\"a}{\"a}},\ and\ \citenamefont {Li}}]{liu2021Optomechanical}%
  \BibitemOpen
  \bibfield  {author} {\bibinfo {author} {\bibfnamefont {Y.}~\bibnamefont {Liu}}, \bibinfo {author} {\bibfnamefont {Q.}~\bibnamefont {Liu}}, \bibinfo {author} {\bibfnamefont {S.}~\bibnamefont {Wang}}, \bibinfo {author} {\bibfnamefont {Z.}~\bibnamefont {Chen}}, \bibinfo {author} {\bibfnamefont {M.~A.}\ \bibnamefont {Sillanp{\"a}{\"a}}},\ and\ \bibinfo {author} {\bibfnamefont {T.}~\bibnamefont {Li}},\ }\bibfield  {title} {\bibinfo {title} {Optomechanical {{Anti-Lasing}} with {{Infinite Group Delay}} at a {{Phase Singularity}}},\ }\href {https://doi.org/10.1103/PhysRevLett.127.273603} {\bibfield  {journal} {\bibinfo  {journal} {Phys. Rev. Lett.}\ }\textbf {\bibinfo {volume} {127}},\ \bibinfo {pages} {273603} (\bibinfo {year} {2021})}\BibitemShut {NoStop}%
\bibitem [{\citenamefont {Liu}\ \emph {et~al.}(2025)\citenamefont {Liu}, \citenamefont {Sun}, \citenamefont {Liu}, \citenamefont {Wu}, \citenamefont {Sillanp{\"a}{\"a}},\ and\ \citenamefont {Li}}]{liu2025Degeneracybreaking}%
  \BibitemOpen
  \bibfield  {author} {\bibinfo {author} {\bibfnamefont {Y.}~\bibnamefont {Liu}}, \bibinfo {author} {\bibfnamefont {H.}~\bibnamefont {Sun}}, \bibinfo {author} {\bibfnamefont {Q.}~\bibnamefont {Liu}}, \bibinfo {author} {\bibfnamefont {H.}~\bibnamefont {Wu}}, \bibinfo {author} {\bibfnamefont {M.~A.}\ \bibnamefont {Sillanp{\"a}{\"a}}},\ and\ \bibinfo {author} {\bibfnamefont {T.}~\bibnamefont {Li}},\ }\bibfield  {title} {\bibinfo {title} {Degeneracy-breaking and long-lived multimode microwave electromechanical systems enabled by cubic silicon-carbide membrane crystals},\ }\href {https://doi.org/10.1038/s41467-025-56497-3} {\bibfield  {journal} {\bibinfo  {journal} {Nat. Commun.}\ }\textbf {\bibinfo {volume} {16}},\ \bibinfo {pages} {1207} (\bibinfo {year} {2025})}\BibitemShut {NoStop}%
\bibitem [{\citenamefont {Teufel}\ \emph {et~al.}(2011)\citenamefont {Teufel}, \citenamefont {Donner}, \citenamefont {Li}, \citenamefont {Harlow}, \citenamefont {Allman}, \citenamefont {Cicak}, \citenamefont {Sirois}, \citenamefont {Whittaker}, \citenamefont {Lehnert},\ and\ \citenamefont {Simmonds}}]{teufel2011Sidebanda}%
  \BibitemOpen
  \bibfield  {author} {\bibinfo {author} {\bibfnamefont {J.~D.}\ \bibnamefont {Teufel}}, \bibinfo {author} {\bibfnamefont {T.}~\bibnamefont {Donner}}, \bibinfo {author} {\bibfnamefont {D.}~\bibnamefont {Li}}, \bibinfo {author} {\bibfnamefont {J.~W.}\ \bibnamefont {Harlow}}, \bibinfo {author} {\bibfnamefont {M.~S.}\ \bibnamefont {Allman}}, \bibinfo {author} {\bibfnamefont {K.}~\bibnamefont {Cicak}}, \bibinfo {author} {\bibfnamefont {A.~J.}\ \bibnamefont {Sirois}}, \bibinfo {author} {\bibfnamefont {J.~D.}\ \bibnamefont {Whittaker}}, \bibinfo {author} {\bibfnamefont {K.~W.}\ \bibnamefont {Lehnert}},\ and\ \bibinfo {author} {\bibfnamefont {R.~W.}\ \bibnamefont {Simmonds}},\ }\bibfield  {title} {\bibinfo {title} {Sideband cooling of micromechanical motion to the quantum ground state},\ }\href {https://doi.org/10.1038/nature10261} {\bibfield  {journal} {\bibinfo  {journal} {Nature}\ }\textbf {\bibinfo {volume} {475}},\ \bibinfo {pages} {359} (\bibinfo {year} {2011})}\BibitemShut {NoStop}%
\bibitem [{\citenamefont {Palomaki}\ \emph {et~al.}(2013)\citenamefont {Palomaki}, \citenamefont {Harlow}, \citenamefont {Teufel}, \citenamefont {Simmonds},\ and\ \citenamefont {Lehnert}}]{palomaki2013Coherent}%
  \BibitemOpen
  \bibfield  {author} {\bibinfo {author} {\bibfnamefont {T.~A.}\ \bibnamefont {Palomaki}}, \bibinfo {author} {\bibfnamefont {J.~W.}\ \bibnamefont {Harlow}}, \bibinfo {author} {\bibfnamefont {J.~D.}\ \bibnamefont {Teufel}}, \bibinfo {author} {\bibfnamefont {R.~W.}\ \bibnamefont {Simmonds}},\ and\ \bibinfo {author} {\bibfnamefont {K.~W.}\ \bibnamefont {Lehnert}},\ }\bibfield  {title} {\bibinfo {title} {Coherent state transfer between itinerant microwave fields and a mechanical oscillator},\ }\href {https://doi.org/10.1038/nature11915} {\bibfield  {journal} {\bibinfo  {journal} {Nature}\ }\textbf {\bibinfo {volume} {495}},\ \bibinfo {pages} {210} (\bibinfo {year} {2013})}\BibitemShut {NoStop}%
\bibitem [{\citenamefont {Liu}\ \emph {et~al.}(2023)\citenamefont {Liu}, \citenamefont {Liu}, \citenamefont {Sun}, \citenamefont {Chen}, \citenamefont {Wang},\ and\ \citenamefont {Li}}]{liu2023Coherenta}%
  \BibitemOpen
  \bibfield  {author} {\bibinfo {author} {\bibfnamefont {Y.}~\bibnamefont {Liu}}, \bibinfo {author} {\bibfnamefont {Q.}~\bibnamefont {Liu}}, \bibinfo {author} {\bibfnamefont {H.}~\bibnamefont {Sun}}, \bibinfo {author} {\bibfnamefont {M.}~\bibnamefont {Chen}}, \bibinfo {author} {\bibfnamefont {S.}~\bibnamefont {Wang}},\ and\ \bibinfo {author} {\bibfnamefont {T.}~\bibnamefont {Li}},\ }\bibfield  {title} {\bibinfo {title} {Coherent memory for microwave photons based on long-lived mechanical excitations},\ }\href {https://doi.org/10.1038/s41534-023-00749-x} {\bibfield  {journal} {\bibinfo  {journal} {npj Quantum Inf.}\ }\textbf {\bibinfo {volume} {9}},\ \bibinfo {pages} {80} (\bibinfo {year} {2023})}\BibitemShut {NoStop}%
\bibitem [{\citenamefont {Bozkurt}\ \emph {et~al.}(2025)\citenamefont {Bozkurt}, \citenamefont {Golami}, \citenamefont {Yu}, \citenamefont {Tian},\ and\ \citenamefont {Mirhosseini}}]{bozkurt2025Mechanical}%
  \BibitemOpen
  \bibfield  {author} {\bibinfo {author} {\bibfnamefont {A.~B.}\ \bibnamefont {Bozkurt}}, \bibinfo {author} {\bibfnamefont {O.}~\bibnamefont {Golami}}, \bibinfo {author} {\bibfnamefont {Y.}~\bibnamefont {Yu}}, \bibinfo {author} {\bibfnamefont {H.}~\bibnamefont {Tian}},\ and\ \bibinfo {author} {\bibfnamefont {M.}~\bibnamefont {Mirhosseini}},\ }\bibfield  {title} {\bibinfo {title} {A mechanical quantum memory for microwave photons},\ }\href {https://doi.org/10.1038/s41567-025-02975-w} {\bibfield  {journal} {\bibinfo  {journal} {Nat. Phys.}\ ,\ \bibinfo {pages} {1}} (\bibinfo {year} {2025})}\BibitemShut {NoStop}%
\bibitem [{\citenamefont {Bukov}\ \emph {et~al.}(2015)\citenamefont {Bukov}, \citenamefont {D'Alessio},\ and\ \citenamefont {Polkovnikov}}]{bukov2015Universal}%
  \BibitemOpen
  \bibfield  {author} {\bibinfo {author} {\bibfnamefont {M.}~\bibnamefont {Bukov}}, \bibinfo {author} {\bibfnamefont {L.}~\bibnamefont {D'Alessio}},\ and\ \bibinfo {author} {\bibfnamefont {A.}~\bibnamefont {Polkovnikov}},\ }\bibfield  {title} {\bibinfo {title} {Universal high-frequency behavior of periodically driven systems: From dynamical stabilization to {{Floquet}} engineering},\ }\href {https://doi.org/10.1080/00018732.2015.1055918} {\bibfield  {journal} {\bibinfo  {journal} {Adv. Phys.}\ }\textbf {\bibinfo {volume} {64}},\ \bibinfo {pages} {139} (\bibinfo {year} {2015})}\BibitemShut {NoStop}%
\bibitem [{wan(2026)}]{wang2026SI}%
  \BibitemOpen
  \bibfield  {title} {\bibinfo {title} {Supplemental material for ``collective coherent perfect absorption in a synthetic photon-phonon lattice'',(i) {{Floquet}} photon-phonon lattice, (ii) collective dynamics and coherent perfect absorption, (iii) microwave optomechanical device and characterization, (iv) sideband cooling of the mechanical mode},\ }\href@noop {} {\  (\bibinfo {year} {2026})}\BibitemShut {NoStop}%
\bibitem [{\citenamefont {Mirhosseini}\ \emph {et~al.}(2020)\citenamefont {Mirhosseini}, \citenamefont {Sipahigil}, \citenamefont {Kalaee},\ and\ \citenamefont {Painter}}]{mirhosseini2020Superconducting}%
  \BibitemOpen
  \bibfield  {author} {\bibinfo {author} {\bibfnamefont {M.}~\bibnamefont {Mirhosseini}}, \bibinfo {author} {\bibfnamefont {A.}~\bibnamefont {Sipahigil}}, \bibinfo {author} {\bibfnamefont {M.}~\bibnamefont {Kalaee}},\ and\ \bibinfo {author} {\bibfnamefont {O.}~\bibnamefont {Painter}},\ }\bibfield  {title} {\bibinfo {title} {Superconducting qubit to optical photon transduction},\ }\href {https://doi.org/10.1038/s41586-020-3038-6} {\bibfield  {journal} {\bibinfo  {journal} {Nature}\ }\textbf {\bibinfo {volume} {588}},\ \bibinfo {pages} {599} (\bibinfo {year} {2020})}\BibitemShut {NoStop}%
\bibitem [{\citenamefont {Rodrigues}\ \emph {et~al.}(2019)\citenamefont {Rodrigues}, \citenamefont {Bothner},\ and\ \citenamefont {Steele}}]{rodrigues2019Coupling}%
  \BibitemOpen
  \bibfield  {author} {\bibinfo {author} {\bibfnamefont {I.~C.}\ \bibnamefont {Rodrigues}}, \bibinfo {author} {\bibfnamefont {D.}~\bibnamefont {Bothner}},\ and\ \bibinfo {author} {\bibfnamefont {G.~A.}\ \bibnamefont {Steele}},\ }\bibfield  {title} {\bibinfo {title} {Coupling microwave photons to a mechanical resonator using quantum interference},\ }\href {https://doi.org/10.1038/s41467-019-12964-2} {\bibfield  {journal} {\bibinfo  {journal} {Nat. Commun.}\ }\textbf {\bibinfo {volume} {10}},\ \bibinfo {pages} {5359} (\bibinfo {year} {2019})}\BibitemShut {NoStop}%
\bibitem [{\citenamefont {Zoepfl}\ \emph {et~al.}(2020)\citenamefont {Zoepfl}, \citenamefont {Juan}, \citenamefont {Schneider},\ and\ \citenamefont {Kirchmair}}]{zoepfl2020SinglePhoton}%
  \BibitemOpen
  \bibfield  {author} {\bibinfo {author} {\bibfnamefont {D.}~\bibnamefont {Zoepfl}}, \bibinfo {author} {\bibfnamefont {M.~L.}\ \bibnamefont {Juan}}, \bibinfo {author} {\bibfnamefont {C.~M.~F.}\ \bibnamefont {Schneider}},\ and\ \bibinfo {author} {\bibfnamefont {G.}~\bibnamefont {Kirchmair}},\ }\bibfield  {title} {\bibinfo {title} {Single-{{Photon Cooling}} in {{Microwave Magnetomechanics}}},\ }\href {https://doi.org/10.1103/PhysRevLett.125.023601} {\bibfield  {journal} {\bibinfo  {journal} {Phys. Rev. Lett.}\ }\textbf {\bibinfo {volume} {125}},\ \bibinfo {pages} {023601} (\bibinfo {year} {2020})}\BibitemShut {NoStop}%
\bibitem [{\citenamefont {Fazio}\ \emph {et~al.}(2025)\citenamefont {Fazio}, \citenamefont {Keeling}, \citenamefont {Mazza},\ and\ \citenamefont {Schir{\`o}}}]{fazio2025ManyBody}%
  \BibitemOpen
  \bibfield  {author} {\bibinfo {author} {\bibfnamefont {R.}~\bibnamefont {Fazio}}, \bibinfo {author} {\bibfnamefont {J.}~\bibnamefont {Keeling}}, \bibinfo {author} {\bibfnamefont {L.}~\bibnamefont {Mazza}},\ and\ \bibinfo {author} {\bibfnamefont {M.}~\bibnamefont {Schir{\`o}}},\ }\href {https://doi.org/10.48550/arXiv.2409.10300} {\bibinfo {title} {Many-{{Body Open Quantum Systems}}}} (\bibinfo {year} {2025}),\ \Eprint {https://arxiv.org/abs/2409.10300} {arXiv:2409.10300 [quant-ph]} \BibitemShut {NoStop}%
\bibitem [{\citenamefont {Diehl}\ \emph {et~al.}(2010)\citenamefont {Diehl}, \citenamefont {Tomadin}, \citenamefont {Micheli}, \citenamefont {Fazio},\ and\ \citenamefont {Zoller}}]{diehl2010Dynamical}%
  \BibitemOpen
  \bibfield  {author} {\bibinfo {author} {\bibfnamefont {S.}~\bibnamefont {Diehl}}, \bibinfo {author} {\bibfnamefont {A.}~\bibnamefont {Tomadin}}, \bibinfo {author} {\bibfnamefont {A.}~\bibnamefont {Micheli}}, \bibinfo {author} {\bibfnamefont {R.}~\bibnamefont {Fazio}},\ and\ \bibinfo {author} {\bibfnamefont {P.}~\bibnamefont {Zoller}},\ }\bibfield  {title} {\bibinfo {title} {Dynamical {{Phase Transitions}} and {{Instabilities}} in {{Open Atomic Many-Body Systems}}},\ }\href {https://doi.org/10.1103/PhysRevLett.105.015702} {\bibfield  {journal} {\bibinfo  {journal} {Phys. Rev. Lett.}\ }\textbf {\bibinfo {volume} {105}},\ \bibinfo {pages} {015702} (\bibinfo {year} {2010})}\BibitemShut {NoStop}%
\bibitem [{\citenamefont {Wang}\ \emph {et~al.}(2026)\citenamefont {Wang}, \citenamefont {Wang}, \citenamefont {{de Jong}}, \citenamefont {{Mercier de L{\'e}pinay}}, \citenamefont {Zhou}, \citenamefont {Sillanp{\"a}{\"a}},\ and\ \citenamefont {Liu}}]{wang2026Pumpthresholdfree}%
  \BibitemOpen
  \bibfield  {author} {\bibinfo {author} {\bibfnamefont {S.}~\bibnamefont {Wang}}, \bibinfo {author} {\bibfnamefont {C.}~\bibnamefont {Wang}}, \bibinfo {author} {\bibfnamefont {M.~H.~J.}\ \bibnamefont {{de Jong}}}, \bibinfo {author} {\bibfnamefont {L.}~\bibnamefont {{Mercier de L{\'e}pinay}}}, \bibinfo {author} {\bibfnamefont {J.}~\bibnamefont {Zhou}}, \bibinfo {author} {\bibfnamefont {M.~A.}\ \bibnamefont {Sillanp{\"a}{\"a}}},\ and\ \bibinfo {author} {\bibfnamefont {Y.}~\bibnamefont {Liu}},\ }\bibfield  {title} {\bibinfo {title} {Pump-threshold-free frequency comb via cavity floquet engineering},\ }\href {https://doi.org/10.1038/s41467-026-72320-z} {\bibfield  {journal} {\bibinfo  {journal} {Nat. Commun.}\ }\textbf {\bibinfo {volume} {17}},\ \bibinfo {pages} {5811} (\bibinfo {year} {2026})}\BibitemShut {NoStop}%
\bibitem [{\citenamefont {Youssefi}\ \emph {et~al.}(2022)\citenamefont {Youssefi}, \citenamefont {Kono}, \citenamefont {Bancora}, \citenamefont {Chegnizadeh}, \citenamefont {Pan}, \citenamefont {Vovk},\ and\ \citenamefont {Kippenberg}}]{youssefi2022Topological}%
  \BibitemOpen
  \bibfield  {author} {\bibinfo {author} {\bibfnamefont {A.}~\bibnamefont {Youssefi}}, \bibinfo {author} {\bibfnamefont {S.}~\bibnamefont {Kono}}, \bibinfo {author} {\bibfnamefont {A.}~\bibnamefont {Bancora}}, \bibinfo {author} {\bibfnamefont {M.}~\bibnamefont {Chegnizadeh}}, \bibinfo {author} {\bibfnamefont {J.}~\bibnamefont {Pan}}, \bibinfo {author} {\bibfnamefont {T.}~\bibnamefont {Vovk}},\ and\ \bibinfo {author} {\bibfnamefont {T.~J.}\ \bibnamefont {Kippenberg}},\ }\bibfield  {title} {\bibinfo {title} {Topological lattices realized in superconducting circuit optomechanics},\ }\href {https://doi.org/10.1038/s41586-022-05367-9} {\bibfield  {journal} {\bibinfo  {journal} {Nature}\ }\textbf {\bibinfo {volume} {612}},\ \bibinfo {pages} {666} (\bibinfo {year} {2022})}\BibitemShut {NoStop}%
\bibitem [{\citenamefont {Kitaev}(2001)}]{kitaev2001unpaired}%
  \BibitemOpen
  \bibfield  {author} {\bibinfo {author} {\bibfnamefont {A.~Y.}\ \bibnamefont {Kitaev}},\ }\bibfield  {title} {\bibinfo {title} {Unpaired {{Majorana}} fermions in quantum wires},\ }\href {https://doi.org/10.1070/1063-7869/44/10S/S29} {\bibfield  {journal} {\bibinfo  {journal} {Phys. Usp}\ }\textbf {\bibinfo {volume} {44}},\ \bibinfo {pages} {131} (\bibinfo {year} {2001})}\BibitemShut {NoStop}%
\bibitem [{\citenamefont {Lindner}\ \emph {et~al.}(2011)\citenamefont {Lindner}, \citenamefont {Refael},\ and\ \citenamefont {Galitski}}]{lindner2011Floquet}%
  \BibitemOpen
  \bibfield  {author} {\bibinfo {author} {\bibfnamefont {N.~H.}\ \bibnamefont {Lindner}}, \bibinfo {author} {\bibfnamefont {G.}~\bibnamefont {Refael}},\ and\ \bibinfo {author} {\bibfnamefont {V.}~\bibnamefont {Galitski}},\ }\bibfield  {title} {\bibinfo {title} {Floquet topological insulator in semiconductor quantum wells},\ }\href {https://doi.org/10.1038/nphys1926} {\bibfield  {journal} {\bibinfo  {journal} {Nat. Phys.}\ }\textbf {\bibinfo {volume} {7}},\ \bibinfo {pages} {490} (\bibinfo {year} {2011})}\BibitemShut {NoStop}%
\bibitem [{\citenamefont {Walther}\ \emph {et~al.}(2006)\citenamefont {Walther}, \citenamefont {Varcoe}, \citenamefont {Englert},\ and\ \citenamefont {Becker}}]{walther2006Cavity}%
  \BibitemOpen
  \bibfield  {author} {\bibinfo {author} {\bibfnamefont {H.}~\bibnamefont {Walther}}, \bibinfo {author} {\bibfnamefont {B.~T.~H.}\ \bibnamefont {Varcoe}}, \bibinfo {author} {\bibfnamefont {B.-G.}\ \bibnamefont {Englert}},\ and\ \bibinfo {author} {\bibfnamefont {T.}~\bibnamefont {Becker}},\ }\bibfield  {title} {\bibinfo {title} {Cavity quantum electrodynamics},\ }\href {https://doi.org/10.1088/0034-4885/69/5/R02} {\bibfield  {journal} {\bibinfo  {journal} {Rep. Prog. Phys.}\ }\textbf {\bibinfo {volume} {69}},\ \bibinfo {pages} {1325} (\bibinfo {year} {2006})}\BibitemShut {NoStop}%
\bibitem [{\citenamefont {Zhao}\ \emph {et~al.}(2021)\citenamefont {Zhao}, \citenamefont {Wu}, \citenamefont {Li}, \citenamefont {Liu}, \citenamefont {Nori}, \citenamefont {Liu},\ and\ \citenamefont {Du}}]{zhao2021PhaseControlled}%
  \BibitemOpen
  \bibfield  {author} {\bibinfo {author} {\bibfnamefont {J.}~\bibnamefont {Zhao}}, \bibinfo {author} {\bibfnamefont {L.}~\bibnamefont {Wu}}, \bibinfo {author} {\bibfnamefont {T.}~\bibnamefont {Li}}, \bibinfo {author} {\bibfnamefont {Y.-x.}\ \bibnamefont {Liu}}, \bibinfo {author} {\bibfnamefont {F.}~\bibnamefont {Nori}}, \bibinfo {author} {\bibfnamefont {Y.}~\bibnamefont {Liu}},\ and\ \bibinfo {author} {\bibfnamefont {J.}~\bibnamefont {Du}},\ }\bibfield  {title} {\bibinfo {title} {Phase-{{Controlled Pathway Interferences}} and {{Switchable Fast-Slow Light}} in a {{Cavity-Magnon Polariton System}}},\ }\href {https://doi.org/10.1103/PhysRevApplied.15.024056} {\bibfield  {journal} {\bibinfo  {journal} {Phys. Rev. Appl.}\ }\textbf {\bibinfo {volume} {15}},\ \bibinfo {pages} {024056} (\bibinfo {year} {2021})}\BibitemShut {NoStop}%
\bibitem [{\citenamefont {Blais}\ \emph {et~al.}(2021)\citenamefont {Blais}, \citenamefont {Grimsmo}, \citenamefont {Girvin},\ and\ \citenamefont {Wallraff}}]{blais2021Circuit}%
  \BibitemOpen
  \bibfield  {author} {\bibinfo {author} {\bibfnamefont {A.}~\bibnamefont {Blais}}, \bibinfo {author} {\bibfnamefont {A.~L.}\ \bibnamefont {Grimsmo}}, \bibinfo {author} {\bibfnamefont {S.~M.}\ \bibnamefont {Girvin}},\ and\ \bibinfo {author} {\bibfnamefont {A.}~\bibnamefont {Wallraff}},\ }\bibfield  {title} {\bibinfo {title} {Circuit quantum electrodynamics},\ }\href {https://doi.org/10.1103/RevModPhys.93.025005} {\bibfield  {journal} {\bibinfo  {journal} {Rev. Mod. Phys.}\ }\textbf {\bibinfo {volume} {93}},\ \bibinfo {pages} {025005} (\bibinfo {year} {2021})}\BibitemShut {NoStop}%
\bibitem [{\citenamefont {{\"O}zdemir}\ \emph {et~al.}(2019)\citenamefont {{\"O}zdemir}, \citenamefont {Rotter}, \citenamefont {Nori},\ and\ \citenamefont {Yang}}]{ozdemir2019Parity}%
  \BibitemOpen
  \bibfield  {author} {\bibinfo {author} {\bibfnamefont {{\c S}.~K.}\ \bibnamefont {{\"O}zdemir}}, \bibinfo {author} {\bibfnamefont {S.}~\bibnamefont {Rotter}}, \bibinfo {author} {\bibfnamefont {F.}~\bibnamefont {Nori}},\ and\ \bibinfo {author} {\bibfnamefont {L.}~\bibnamefont {Yang}},\ }\bibfield  {title} {\bibinfo {title} {Parity--time symmetry and exceptional points in photonics},\ }\href {https://doi.org/10.1038/s41563-019-0304-9} {\bibfield  {journal} {\bibinfo  {journal} {Nat. Mater.}\ }\textbf {\bibinfo {volume} {18}},\ \bibinfo {pages} {783} (\bibinfo {year} {2019})}\BibitemShut {NoStop}%
\end{thebibliography}%
\end{document}